\documentclass[12pt,preprint]{aastex}
\usepackage{natbib}

\newcommand{\etal}{et al.}

\newcommand{\msun}{M$_{\sun}$\,}
\newcommand{\myr}{M$_\odot$~yr$^{-1}$} 

\newcommand{\ha}{H$\alpha$}
\newcommand{\hb}{H$\beta$}
\newcommand{\nii}{[N{\sc ii}]}
\newcommand{\oiii}{[O{\sc iii}]}
\newcommand{\hii}{H{\sc ii}}
\newcommand{\kms}{km\,s$^{-1}$}

\shorttitle{Resolved Spectroscopy of z$\sim$1.6 Galaxies}

\shortauthors{Wright, S.~A. \etal~2008}

\begin{document}

\title{Dynamics of Galactic Disks and Mergers at $\MakeLowercase{z}$$\sim$1.6: Spatially Resolved Spectroscopy with Keck Laser Guide Star Adaptive Optics}

\author{Shelley A. Wright\altaffilmark{1},  James E. Larkin\altaffilmark{2}, David R. Law\altaffilmark{2}, Charles C. Steidel\altaffilmark{3}, Alice E. Shapley\altaffilmark{2}, Dawn K. Erb\altaffilmark{4}}

\begin{abstract}

We present 0$\farcs$2 resolution near-infrared integral field spectroscopy of \ha~emission from six star forming galaxies at z$\sim$1.6 (look-back time of $\sim$9.6 Gyr). These observations were obtained with OSIRIS using the Keck Laser Guide Star Adaptive Optics system. All sources have a compact spatial extent of $\sim$1$\arcsec$, with an average half light radius of $\overline{r}$$_{1/2}$=2.9 kpc and average dereddened star formation rate of 22 \myr. Based on \ha~kinematics we find that these six galaxies are dynamically distinguishable, and we classify them as either merger or disk candidate systems. We find three merger systems (HDF-BX1287, HDF-BX1315, and Q1623-BX491) with varying geometries and dynamical properties. Three galaxies (HDF-BMZ1299, Q2343-BX344, and Q2343-BM145) are well-fit by an inclined-disk model with low velocity residuals (20 to 46 \kms). An average plateau velocity of $\overline{v_{p}}$=185 \kms~is achieved within 1.0 kpc. The majority of observed velocity dispersions ($\overline{\sigma}$ $\sim$ 88 \kms) can be explained by the residual seeing halo, and are not intrinsic to our sources. However, one merger and one disk candidate have high velocity dispersions ($\sigma$$_{\rm{obs}}$ $\gtrsim$ 200 \kms) that cannot be solely explained by beam smearing. For two disk candidates, we detect \nii~emission and are able to map the \nii/\ha~ratio on kiloparsec scales. In both cases, \nii~emission is more concentrated than \ha~emission ($\lesssim$0\farcs2), and peak ratios are best explained by the presence of an AGN. These are among the weakest known AGN at high redshift, however their emission is strong enough to impact high redshift metallicity studies that use nebular ratios. All disk candidates have likely completed only a few orbital periods, and if left unperturbed are excellent candidates to become present-day spiral galaxies.

\end{abstract}


\altaffiltext{1}{Department of Physics and Astronomy,
University of California, Irvine, CA 92697;
saw@uci.edu}
\altaffiltext{2}{Department of Physics and Astronomy,
University of California, Los Angeles, CA 90095; 
larkin, drlaw, aes@astro.ucla.edu}
\altaffiltext{3}{California Institute of Technology, 
MS 105-24, Pasadena, CA 91125; 
ccs@astro.caltech.edu}
\altaffiltext{4}{Department of Physics,
University of California, Santa Barbara, CA  93106;
dawn@physics.ucsb.edu}

\section{Introduction}\label{intro}

During the last decade, innovative large surveys have discovered diverse sets of galaxies within the early universe. These extragalactic campaigns have studied the global properties of high-redshift (z $\gtrsim$ 1)  galaxies, such as their masses, stellar populations, star formation rates, metallicity, and dynamics  (eg, \citealt{madau96, steidel03, steidel04, van04, pap05, erb06a,erb06b}). Yet even with rich data sets we still struggle to understand the evolution of these galaxies into the modern Hubble sequence.

Recent advances of diffraction-limited techniques on 8-10m telescopes using adaptive optics (AO) and integral field spectroscopy (IFS) have greatly enhanced dynamical studies of early galaxies. An IFS coupled with an AO system provides a unique ability to probe the spatial extent of high-redshift star forming galaxies and map their internal dynamics at sub-kiloparsec scales. In the last three years, near-infrared IFS, such as OSIRIS \citep{larkin06} at Keck Observatory and SINFONI \citep{eisen03} at VLT, have offered a new and more intimate view of the complicated processes of galaxy formation. These studies are challenging, since IFS observations can only examine a small sample of the brightest galaxies due to the large amount of observing time required to achieve a significant signal-to-noise (S/N). The current samples of 3D spectroscopy \citep{schreib06,genzel06,wright07,law07, bou07, genzel08} show there are a wide variety of kinematic and morphological states of early galaxies. It is still unclear whether the majority of star formation occurs in flattened disk systems as anticipated by some models \citep{mo98}, or disordered configurations of merger systems, or some other unrecognized geometry.

The first results with near-infrared IFS have found candidate disk galaxies at look back times comparable to or greater than the age of the Milky Way disk, but there are significant unexplained differences in gas kinematics compared to local galaxies \citep{schreib06,genzel06,wright07,law07,genzel08}. In particular, observed velocity dispersions seen in early disk candidates are much higher than in local spiral galaxies, which likely reflects a high level of turbulence within these systems. In addition, the actual age of the Milky Way disk remains controversial and thus does not unambiguously point to an era of disk formation. Detailed modeling of white dwarf luminosities has yielded a Milky Way disk age of 7.3$\pm$1.5 Gyr, which indicates a significant gap in time between the formation of the halo and the disk \citep{hans02}. However, there is the possibility that studies have missed a population of faint disk white dwarfs, which would then yield an older Milky Way disk age. Isochrone fitting of disk clusters show that the cluster NGC 6791 (a member of the thick-disk) has an age between 7.5 and 10 Gyr, with the uncertainty mostly due to the allowed range of metallicity \citep{sand03}. With the most commonly accepted cosmology \citep{benn03}, a lookback time of 7.5 Gyr corresponds to z$\sim$1.0 while 10 Gyr is close to z$\sim$2.0. Therefore galaxies close to the look back time of the estimated ages of the oldest stars within the Milky Way disk are candidates for discovering early disk systems. 

From the theoretical side, detailed simulations show that an anti-hierarchical scenario is possible within a $\Lambda$-CDM universe \citep{springel06}. In these simulations it has been recognized that more massive systems arise in regions of greatest over-density, and thus gain a head start in their formation. Therefore more massive galaxies have resulted from a larger number of mergers, and thus evolve faster than lower mass systems. In addition, semi-analytic models \citep{somer01,somer06} that include energy feedback from star formation are able to prevent baryons from losing the majority of their angular momentum during the initial collapse phase, thus allowing large mature disks to survive. Although we expect the early rate of star formation to be extremely high (50-100 \myr), it is unknown what the spatial distribution, temporal duty cycle, internal turbulence and global kinematics of star formation are within these early disks. This leaves the form of energy released into the system to prevent cooling largely a free parameter. Thus, the discovery of early disks and measurement of their rotational speeds, internal velocity dispersions and sizes are important in constraining models of galaxy formation, and in tying together high redshift objects with their modern counterparts.

We have conducted a Laser Guide Star Adaptive Optics (LGS-AO) survey of z$\sim$1.6 star forming galaxies using the IFS OSIRIS at Keck Observatory. In this paper we present two-dimensional kinematics of six z$\sim$1.6 star forming galaxies, with varying morphologies and dynamics of both merger and disk candidate systems. In \S\ref{observ} we describe the observations and reductions. In \S\ref{sed}, we present results for stellar population fits to each of the galaxy's Spectral Energy Distributions (SED) from previous multi-wavelength imaging studies. In \S\ref{spectral} we present our analysis, with flux distribution, star formation rates, and metallicity of each source. We discuss results and analysis in \S\ref{dyn} of \ha~and \nii~kinematics, disk modeling, beam smearing, and mass estimates. Previous observations of optical spectroscopy are investigated for their kinematic information on rest-frame UV absorption and emission features in \S\ref{uv}. Each of our galaxies have unique characteristics, which are discussed on an individual basis in \S\ref{indiv}. In \S\ref{discuss} we compare our study to other larger surveys of star forming galaxies at a similar epoch, and to present-day systems. Throughout the paper we assume $\Lambda$-dominated cosmology with $\Omega_{M}$=0.27, $\Omega_{V}$=0.73, and H$_{o}$=71 \kms~ Mpc$^{-1}$ \citep{benn03}.
 
\section{OSIRIS Observations \& Data Reduction}\label{observ}

Observations of six star forming galaxies (SFGs) between z = 1.5 - 1.7 were obtained at the W.M. Keck II 10m telescope using OSIRIS with the LGS-AO system \citep{wiz06,mvdam06}. OSIRIS is a near-infrared integral field spectrograph which uses a lenslet array to obtain up to 3,000 simultaneous spectra over a rectangular field of view. We used the coarsest spatial scale of 0\farcs1 with the corresponding narrowband filter which contains the redshifted \ha~and \nii~emission. This mode yields a spatial sampling of 0\farcs1 in both lenslet axes with a 4\farcs8 x 6\farcs4 field of view at spectral resolution of R$\sim$ 3400. We selected our targets from the rest-frame UV color selected catalog of \citet{steidel04} and when available used information from previous near-infrared spectroscopic observations from \citet{erb06a}. When previous \ha~observations were unavailable, objects whose luminosities and colors indicated high star formation rates were prioritized.
 
For all six SFGs the LGS-AO system produced an artifical guide star directly on the OSIRIS optical axis, with the tip-tilt sensor locked onto an off-axis natural guide star. A position angle (PA) was selected to maintain both the galaxy and tip-tilt star in the OSIRIS and LGS-AO fields of view, respectively. A Strehl ratio was estimated for each SFG observation given the magnitude and separation of the tip-tilt star based on the statistical performance of the Keck AO system. To ensure acquisition we observed the tip-tilt star with OSIRIS LGS-AO 0\farcs1 lenslet scale before moving to the galaxy. For some of the galaxies we observed these tip-tilt stars with the finer spatial scale of 0\farcs02 to assess the AO performance and measured the point spread function full-width at half-maximum (FWHM) to be $\lesssim$ 0\farcs1. For tip-tilt stars solely observed in the coarse 0\farcs1 scsale, we have undersampled the PSF of the AO correction, so this represents an upper limit to the achieved performance. A series of individual 900 second frames were taken on source dithered 1\arcsec~north and south of the field center, with at least one 900 second sky exposure 5\arcsec off the science target for optimal sky subtraction. The total exposure time for each of the sources was 1-1.5 hours. Table \ref{obs} lists the observational details using OSIRIS and the LGS-AO system for each of the SFGs. Throughout the paper, when discussing multiple targets we list their order according to increasing right ascension. 

Data reduction was performed using the OSIRIS final data reduction pipeline (DRP) in combination with  custom IDL routines.  Typical near-infrared reduction steps were carried out using the DRP: correction of detector non-linearity, removal of bad pixels, removal of detector cross talk, sky subtraction, wavelength calibration, flux calibration, and mosaicking of individual frames. Critical and unique reduction steps for this data set are extraction of emission line spectra and assembly of the 3D cube (x and y spatial locations of lenslets and $\lambda$ wavelength), in which the flux for each lenslet is the integral of the peak pixel values for each wavelength channel. Each observing night we observed an Elias standard \citep{elias82} and flux calibrated the standard per wavelength channel. In addition, to optimize subtraction of sky OH lines from the object we have written custom IDL software to perform a scaled sky subtraction using atmospheric OH-emission lines near the observed  \ha~emission.

\section{Spectral Energy Distibutions: Ages, Extinction, and Stellar Mass}\label{sed}

Our galaxies were selected by their optical colors in U$_{n}$GR wavebands \citep{steidel04}, and were all confirmed with optical spectroscopy using LRIS on the Keck I telescope. The Steidel et al. survey primarily focused on  z$\sim$2-3 galaxies, but has also used the two-color technique outlined in \citet{adel04} to detect lower redshift z$\sim$1.6 star-forming counterparts within the traditional redshift desert \citep{steidel04}. Table \ref{photo} presents the multi-wavelength photometric data for each of the galaxies. Near-infrared observations in J and K$_{s}$ were taken with the WIRC camera on Palomar (5m) and are described in \citet{erb06b}. Mid-infrared observations were taken with Spitzer IRAC (3.6, 4.5, 5.8, and 8.5 \micron) and MIPS (24 \micron) for all galaxies, and observations and magnitudes of the HDF galaxies are outlined in \citet{reddy06} . Best-fit models of the stellar population were found for each galaxy's SED from observed $\sim$0.35$\sbond$8.5 \micron~photometry. The method of determining the best-fit SED is fully discussed in \citet{shap01,shap05}, and more recently in \citet{erb06a,erb06b} for their z$\sim$2 sample. For this fitting analysis, a grid of a solar metallicity \citep{bruz03} were used with a \citet{chab03} Initial Mass Function (IMF) for a \citet{calz00} starburst attenuation law with a range of ages, extinction, and constant star formation histories \citep{erb06b}. We present best-fit ages, E(B-V) extinction, and total stellar mass for each galaxy in Table \ref{photo}.

\section{OSIRIS Spectral Properties}\label{spectral}

\begin{figure*}[t]
\epsscale{0.6}
\begin{center}
\plotone{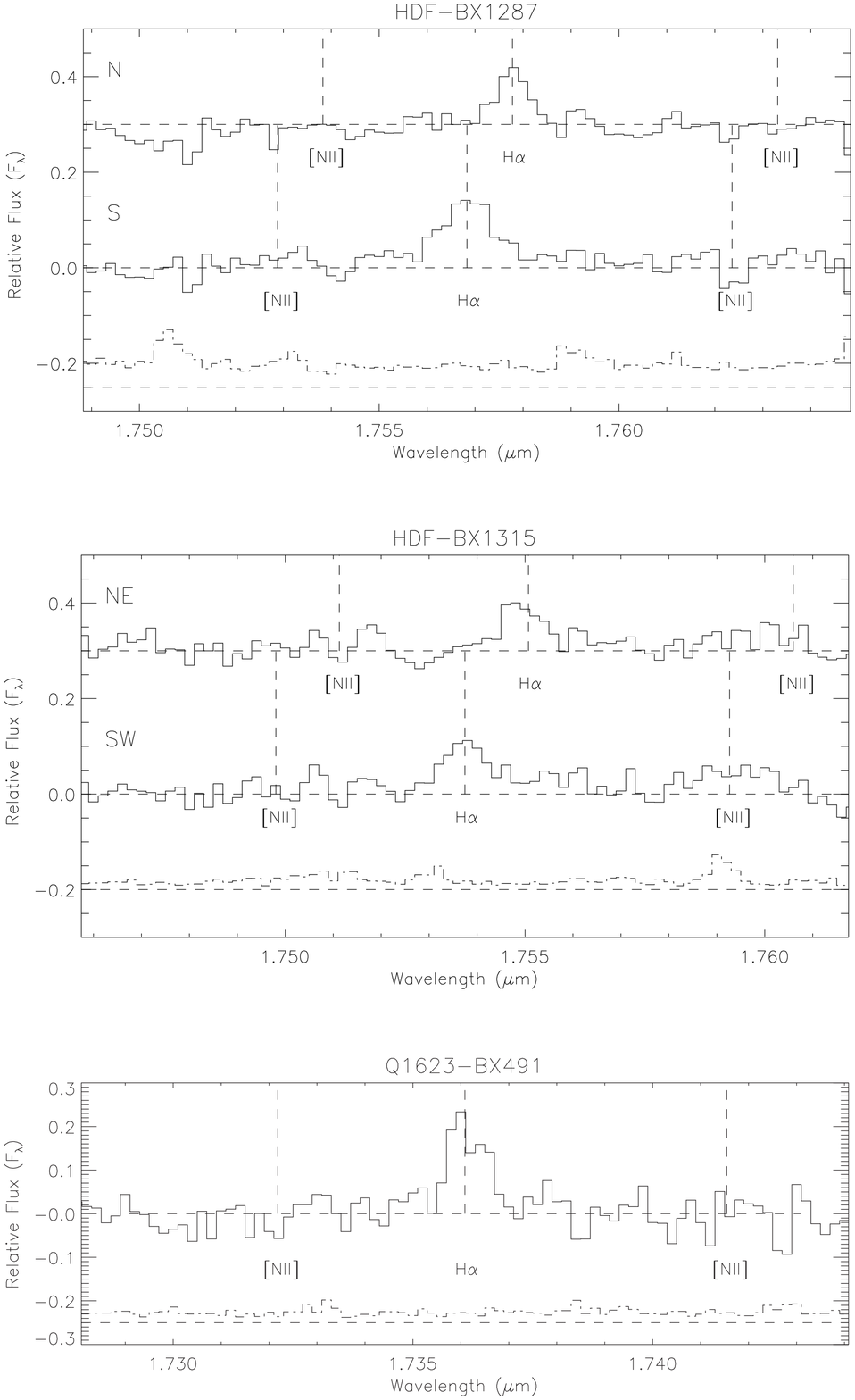}
\caption{\linespread{1}\small\textit{ 
Integrated \ha~spectra (F$_{\lambda}$) of merger candidates HDF-BX1287, HDF-BX1315, and Q1623-BX491. There are two spectra for HDF-BX1287 (N and S) and HDF-BX1315 (NE and SW), collapsed over a 0\farcs5x0\farcs5 spatial region on each of the two resolved knots for each galaxy. The spectrum for Q1623-BX491 is collapsed spatially over a 1\farcs0x1\farcs0 region of the galaxy. Noise per spectral channel is plotted to illustrate the locations of OH night sky emission (dot-dash line). \ha~is marked and labeled for each source, as are expected locations of \nii~emission based on the \ha~centroid. 
  }}\label{merger_spec}
 \end{center}
\end{figure*}

\begin{figure*}[t]
\epsscale{0.6}
\begin{center}
\plotone{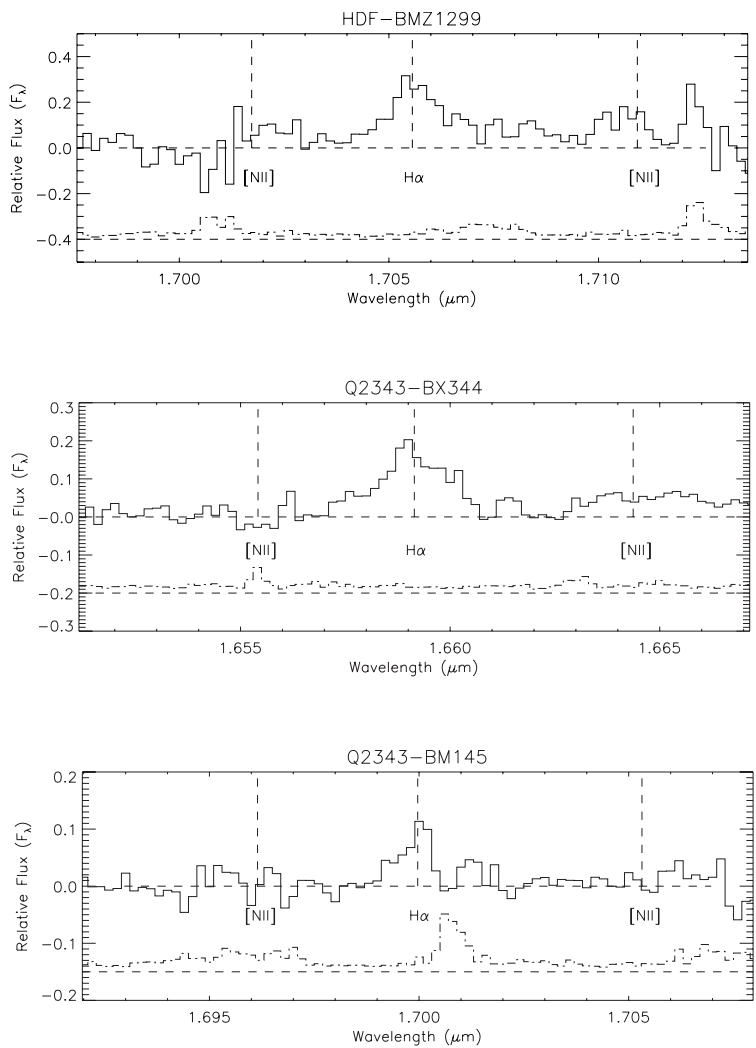}
\caption{\linespread{1}\small\textit{ 
Integrated \ha~spectra (F$_{\lambda}$) for disk candidates HDF-BMZ1299, Q2343-BX344, and Q2343-BM145 spatially collapsed over 1\farcs0x1\farcs0 region of the galaxy. Noise per spectral channel is plotted to illustrate locations of OH night sky emission (dot-dash line). \ha~is marked and labeled for each source, as are expected locations of \nii~emission based on the \ha~centroid. 
  }}\label{disk_spec}
\end{center}
\end{figure*}

A spatially integrated spectrum for each source was generated by collapsing over the entire spatial extent of the object to derive the systemic redshift (z$_{H_{\alpha}}$) and global velocity dispersion ($\sigma$$_{\rm{global}}$). The total \ha~flux was determined from the summation of the flux over the entire spatial and spectral extent of the object. Tables \ref{global} and \ref{kinemass} list global flux and kinematic properties of the sources. Figures \ref{merger_spec} and \ref{disk_spec} show the global spectra for the merger and disk candidates, respectively. The central peak of \ha~emission within the global spectrum is marked, and the anticipated location of both \nii~emission lines are indicated (6550$\AA$ and 6585$\AA$). For two of the sources,  HDF-BX1287 and HDF-BX1315, global spectra for each resolved, separate knot are presented. Included underneath each object's spectrum is the standard deviation of the sky plotted per wavelength channel to illustrate noisy locations where OH sky lines are positioned. 

\subsection{\ha~Flux Distribution}\label{flux}

\begin{figure*}[t]
\epsscale{0.8}
\begin{center}
\plotone{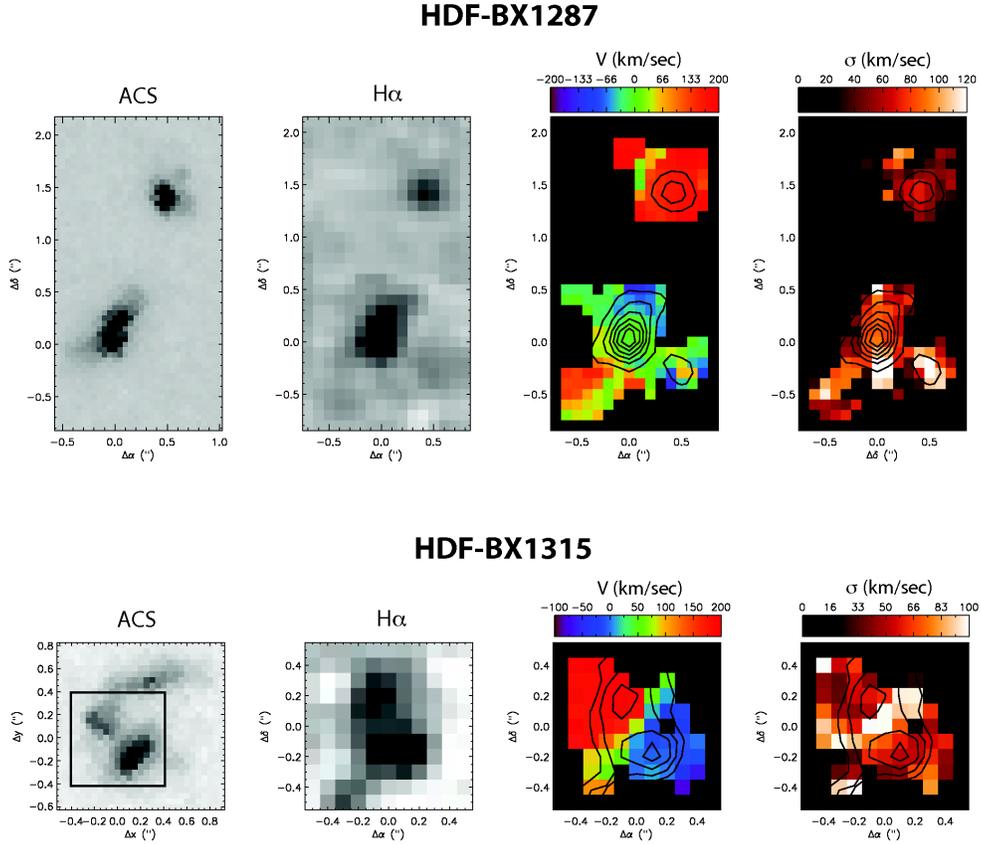}
\caption{\linespread{1}\small\textit{ 
Two-dimensional \ha~flux distribution and kinematics of merger candidates HDF-BX1287 and HDF-BX1315.  (Left) HST ACS color stacked image (B,V, I, Z). (Middle Left) Integrated \ha~flux observed with OSIRIS. For HDF-BX1287, two different wavelength regimes were collapsed and combined to make the final integrated \ha~image. (Middle Right) Two-dimensional \ha~kinematics showing spatial distribution of velocity centers (km s$^{-1}$). (Right) Two-dimensional kinematics showing the velocity dispersion (km s$^{-1}$) map.  }}\label{acs}
\end{center}
\end{figure*}

\begin{figure*}[t]
\epsscale{0.8}
\begin{center}
\plotone{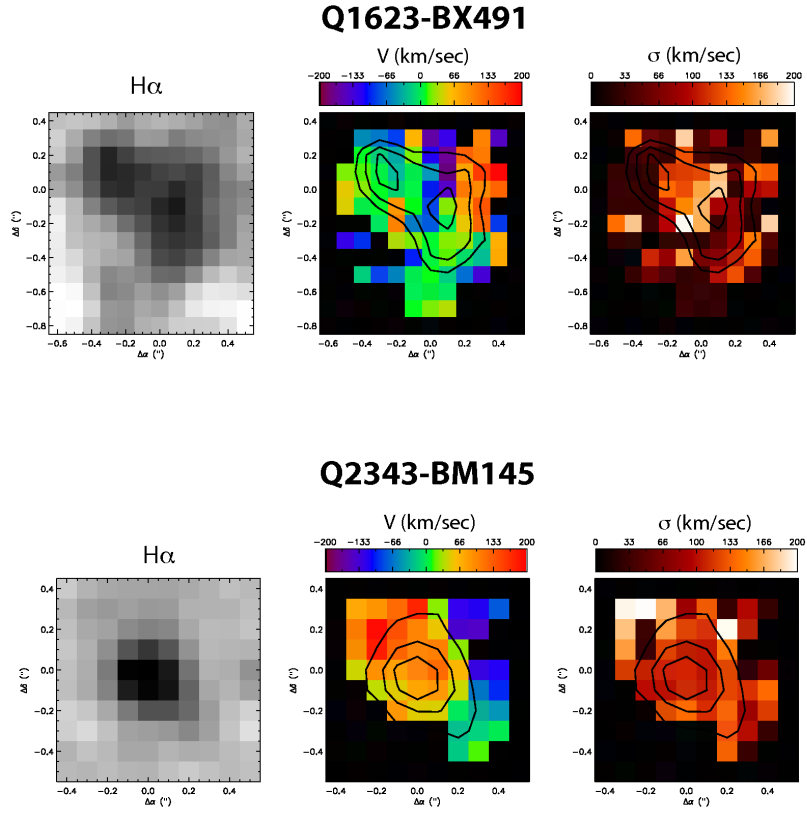}
\caption{\linespread{1}\small\textit{ 
Two-dimensional \ha~flux distribution and kinematics of merger candidate Q1623-BX491 and disk candidate Q2343-BM145.(Left) Integrated \ha~flux observed from OSIRIS. (Middle) Two-dimensional \ha~kinematics showing spatial distribution of velocity centers (km s$^{-1}$). (Right) Two-dimensional kinematics showing the velocity dispersion (km s$^{-1}$) map.
  }}\label{twoobj}
\end{center}
\end{figure*}

Integrated \ha~flux images for each SFG are generated by collapsing over spectral channels which contain the peak emission from the source. To achieve higher signal-to-noise ratio (S/N) for kinematic analysis we spatially smoothed each final mosaicked image per wavelength channel with a Gaussian kernel (FWHM = 0\farcs2). Figures \ref{acs},  \ref{twoobj}, \ref{bmz1299_fig}, and \ref{bx344_fig} contain images of the \ha~flux distributions for each of the galaxies. Within the figures, centers of the images are set at the peak \ha~flux spatial location. Five of the sources are resolved over $\sim$ 1$\arcsec$ (8.5 kpc) in size, with three resolved sources having two distinct knot-like structures (Q1623-BX491, HDF-BX1315, and HDF-BMZ1299) and two that consist of a single compact source (Q2343-BX344 and Q2343-BM145). The sixth source, HDF-BX1287, is a multiple knot system separated at its greatest distance by $\sim$ 2.5$\arcsec$ (21.4 kpc).

\subsection{Star Formation Rates}\label{sfr}

\ha~emission traces the formation of massive stars through hydrogen recombination in \hii~regions, and provides an instantaneous measurement of star formation at the observed epoch, but not necessarily the underlying stellar population and intrinsic morphology. Star formation rates are derived for each source using the global Schmidt law [SFR (\msun yr$^{-1}$) = L$_{H\alpha}$ / 2.23 x 10$^{41}$ ergs s$^{-1}$; \citet{kenn98} with a \citet{chab03} IMF]. \ha~luminosities are found by using the total flux, systemic redshift, and concordance cosmology and are listed in Table \ref{global}. To account for extinction we use the E(B-V) previously derived from modeling the SED (\S\ref{sed}) for these sources to find a dereddened flux and hence SFR. The dust corrected SFR for these sources are between $\sim$5$\sbond$33 \msun yr$^{-1}$. We determine an \ha~surface brightness limit for each source that has \ha~emission with a S/N$\gtrsim$5, and find a range of 0.6$\sbond$1.4$\times$10$^{-18}$ ergs s$^{-1}$ cm$^{-2}$ kpc$^{-2}$. These \ha~flux limits correspond to an observed SFR per unit area ($\Sigma_{SFR}$) limit of  $\sim$0.06$\sbond$0.11 \msun yr$^{-1}$ kpc$^{-1}$. Winds from supernova and star formation activity are known to be associated with local star forming galaxies with $\Sigma_{SFR}$ $\gtrsim$0.1\msun yr$^{-1}$ kpc$^{-1}$ \citep{heck02}. Even though the seeing halo may bias  \ha~flux in the outer, lower S/N regions, it is still likely that these galaxies harbor winds across the majority extent of observed \ha~emission. It is interesting to note that all sources exhibit similar SFR even though their kinematics indicate different morphological classification, as we will discuss in \S\ref{dyn}. However, there is a selection bias using OSIRIS LGS-AO for detecting sources that are compact ($\lesssim$ 1\arcsec) and which have a high SFR.
 
\subsection{Metallicity}\label{metal}

Abundances in \hii~regions are traditionally determined by ratios of optical nebular lines. \citet{pett04} calibrated line ratios of \oiii/\hb~and \nii/\ha~to determine metallicities from \hii~regions in high-redshift galaxies. Multiple line indicators has the benefit of distinguishing between star forming galaxies and galaxies with high contributions of AGN activity using traditional \citet{bald81} diagrams. \citet{shap04} used near infrared spectroscopy of seven sources for z $\gtrsim$ 2 galaxies and used the \nii/\ha~line ratio to find at-least solar metallicities at a look-back time of 10.5 Gyr. More recently, \citet{erb06a} extended this study to a larger \nii/\ha~sample and found a mass-metallicity relation for z$\sim$2 galaxies, where metallicity increases with decreasing gas fractions and increasing stellar mass. At lower redshifts, \citet{shap05} targeted a small sample of z$\sim$1.3 galaxies using \oiii/\hb~and \nii/\ha~line ratios, and found a half to two-thirds solar metallicity. However, emission-line ratios found at z$\sim$1.3 \citep{shap05, liu08} are offset upwards when compared to \oiii~and \nii~emission from typical excitation levels found in local galaxies within a \citet{bald81} diagram, which implies either that the physical conditions of star formation at higher redshift may be different, or there are contributions from weak AGN activity.

For two galaxies, HDF-BMZ1299 and Q2343-BX344, we directly detect \nii($\lambda$6585) emission and are able to evaluate global oxygen abundances in the following way. Total \nii~flux is determined by integrating spatially and spectrally at the location of \nii~emission within each source. The nebular line indicator \nii/\ha~is determined by the ratio of the \nii~spatial locations matched to the same location of \ha~emission. For the other four galaxies with no direct \nii~detection, we determine an upper 3$\sigma$ limit for the \nii/\ha~ratio by integrating an 0\farcs5$\times$0\farcs5 aperture box over the predicted \nii~spectral location based on the central peak of \ha~flux (see Figures \ref{merger_spec} and \ref{disk_spec}) . Assuming these galaxies have no AGN contribution, we use \citet{pett04} to infer a metallicity from the measured linear relationship between oxygen abundance and the \nii/\ha~ratio, where a value above 8.66 is considered to exceed a solar metallicity \citep{all01}. 

\begin{equation}
12 + log\left(\frac{O}{H}\right) = 8.90 + 0.57 \times log\left(\frac{[N_{II}]}{H_{\alpha}}\right)
\end{equation}

Within Table \ref{global}, upper limits, directly measured \nii/\ha~ratios and oxygen abundances are listed.  Upper 3$\sigma$ \nii/\ha~limits suggest that these sources have below solar metallicites. Given the compact \nii~emission with high \nii/\ha~ratios, we discuss the likelihood of AGN activity within our two direct \nii~detections, HDF-BMZ1229 and Q2343-BX344, in \S\ref{bmz1299}, \S\ref{bx344} and \S\ref{discuss}.

\section{Dynamical Properties}\label{dyn}

\subsection{\ha~and \nii~Kinematics}\label{kine}

\begin{figure*}[t]
\epsscale{1.0}
\begin{center}
\plotone{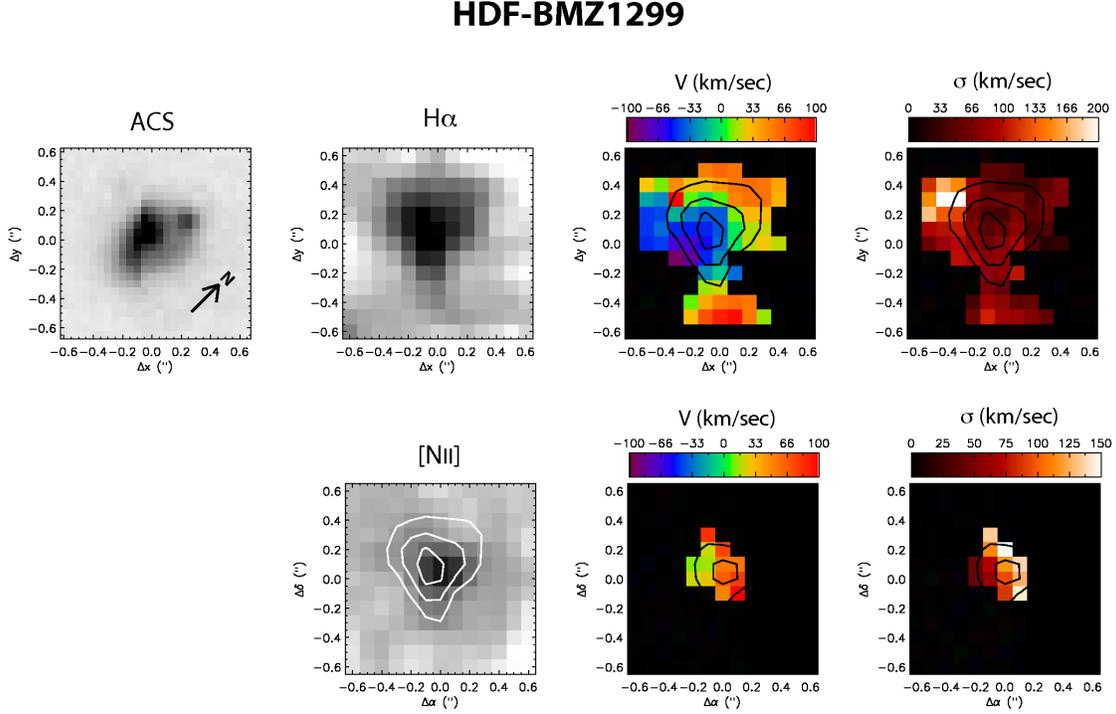}
\caption{\linespread{1}\small\textit{ 
\ha~flux distibution and kinematics for HDF-BMZ1299. All images in the figure have a position angle of 45$^{\circ}$, represented by the arrow in the ACS image. TOP PANEL: (Left) HST ACS color stacked image (B,V,R,I). (Middle Left) Integrated \ha~flux observed with OSIRIS. (Middle Right) Two-dimensional \ha~kinematics showing spatial distribution of velocity centers (km s$^{-1}$). (Right) Two-dimensional kinematics showing the velocity dispersion (km s$^{-1}$) map. Note that the dispersion peak is not coincident with the flux center, consistent with a disk.
BOTTOM PANEL:  \nii~flux distibution and kinematics for HDF-BMZ1299. (Left) Integrated \nii~flux observed with OSIRIS. (Middle) Two-dimensional \nii~kinematics showing spatial distribution of velocity centers (km s$^{-1}$). (Right) Two-dimensional \nii~kinematics showing the velocity dispersion (km s$^{-1}$) map. 
  }}\label{bmz1299_fig}
\end{center}
\end{figure*}

\begin{figure*}[t]
\epsscale{0.8}
\begin{center}
\plotone{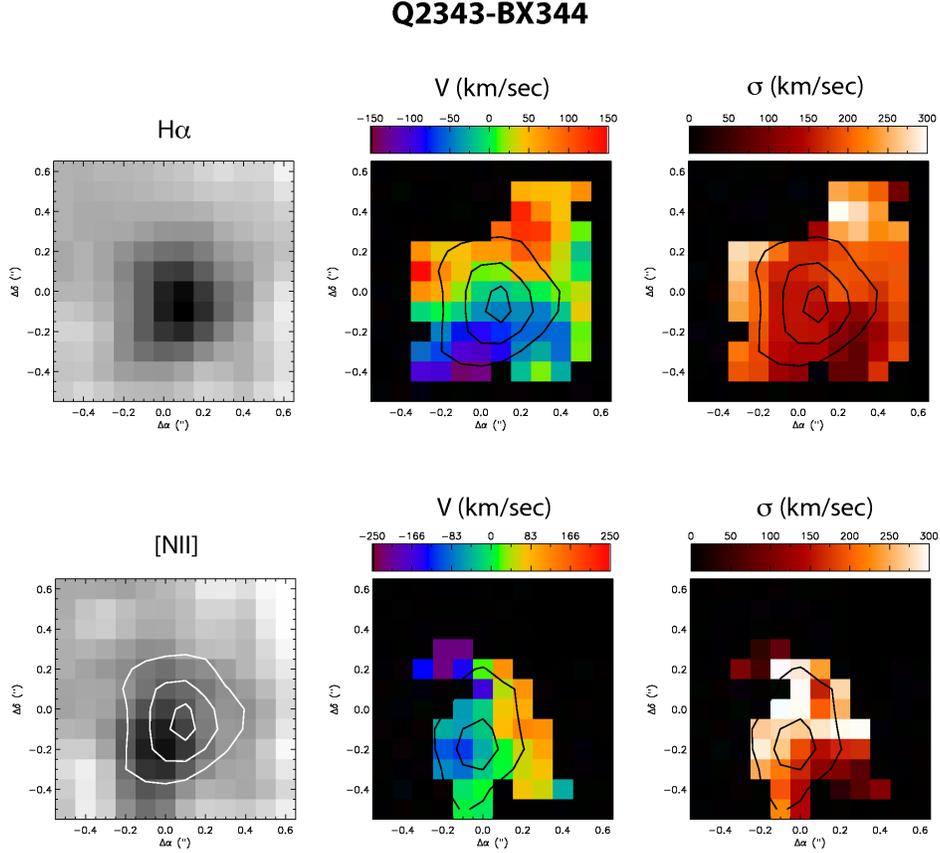}
\caption{\linespread{1}\small\textit{ 
TOP PANEL: \ha~flux distibution and kinematics for Q2343-BX344. (Left) Integrated \ha~flux observed with OSIRIS. (Middle) Two-dimensional \ha~kinematics showing spatial distribution of velocity centers (km s$^{-1}$). (Right) Two-dimensional kinematics showing the velocity dispersion (km s$^{-1}$) map. Note that the peak of the dispersion is not coincident with the flux center, consistent with a disk.
BOTTOM PANEL:  \nii~flux distibution and kinematics for Q2343-BX344. (Left) Integrated \nii~flux observed with OSIRIS, overlapped with \ha~contours. (Middle) Two-dimensional \nii~kinematics showing spatial distribution of velocity centers (km s$^{-1}$). (Right) Two-dimensional \nii~kinematics showing spatial distribution of velocity dispersion (km s$^{-1}$).  }}\label{bx344_fig}
\end{center}
\end{figure*}

Two dimensional \ha~emission velocity maps are generated for each galaxy by fitting a Gaussian to the spectrum at each lenslet with an integrated signal-to-noise (S/N) $\gtrsim$ 5 (each visually verified), yielding both velocity offsets and dispersions corrected for instrumental widths. Velocity offsets are relative to the nominal central wavelength determined from the global Gaussian fit for the entire spatially integrated  spectrum of the galaxy discussed in \S\ref{spectral}. For each lenslet, the Gaussian fit is weighted by the inverse variance per wavelength channel, which is particularly important for galaxies which fall close to atmospheric OH emission lines. Average velocity uncertainties for each source are listed in Table \ref{kinemass}, and are found to range between 5 to 20 \kms~across the galaxies. Figures \ref{acs},  \ref{twoobj}, \ref{bmz1299_fig}, and \ref{bx344_fig} present two dimensional radial velocity and velocity dispersion maps for each galaxy. Three galaxies were observed in HDF-N with the Hubble Advanced Camera System (ACS)\footnotemark, and their color-stacked (B,V,I,Z) images are included within these figures. There are a range of velocity offsets and dispersions, where the highest dispersions are of order $\sigma_{\rm{obs}}$ $\sim$ 200 \kms, and the lowest $\sigma_{\rm{obs}}$ $\sim$ 45 \kms.

\footnotetext{HDF-N ACS images were observed November 21, 2002 with proposal identifier 9583 and principal investigator Giavalisco, and see \citet{law07} for further description}

\begin{figure*}[t]
\epsscale{0.9}
\begin{center}
\plotone{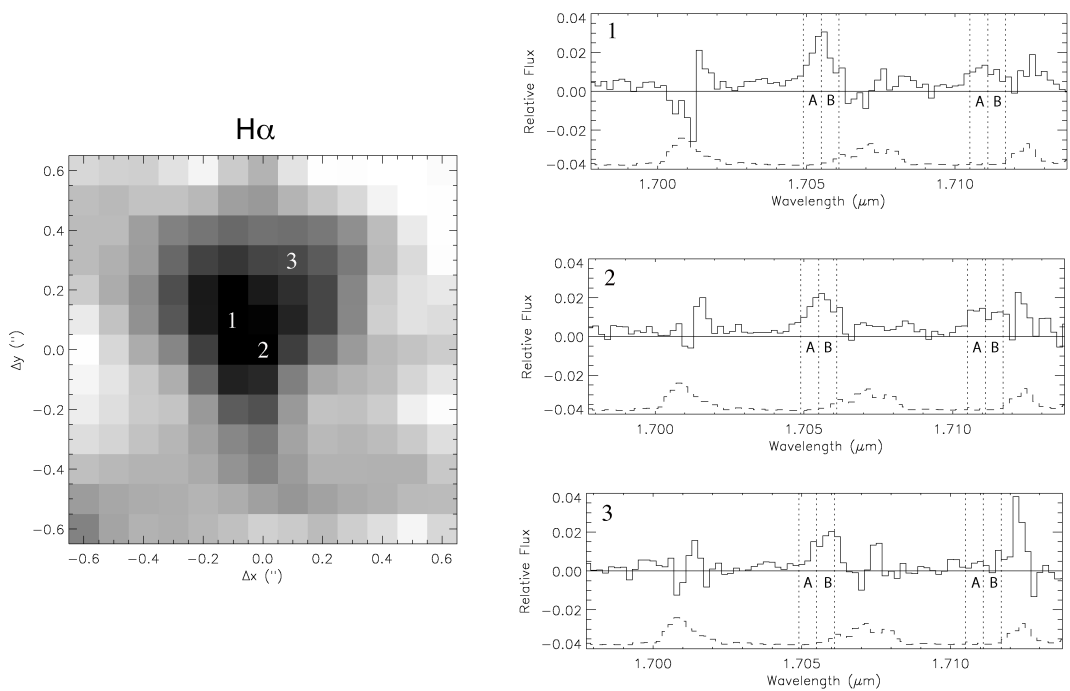}
\caption{\linespread{1}\small\textit{ 
\ha~image of HDF-BMZ1299 with individual lenslet spectra plotted (F$_{\lambda}$) for three different regions of the galaxy labeled 1, 2, and 3. The dashed line represents the standard deviation per spectral channel per lenslet off source. The region labeled `A' represents the \nii/\ha~ratio of a velocity location of \ha~having 0 \kms, and the `B' region represents the \nii/\ha~ratio for a \ha~velocity location of 140 \kms. We present 2D \nii/\ha~ratio maps for each of these identified (A and B) spectral locations in Figure \ref{metalmaps}. 
}}\label{bmz1299_indiv_fig}
\end{center}
\end{figure*}

\begin{figure*}[t]
\epsscale{0.9}
\begin{center}
\plotone{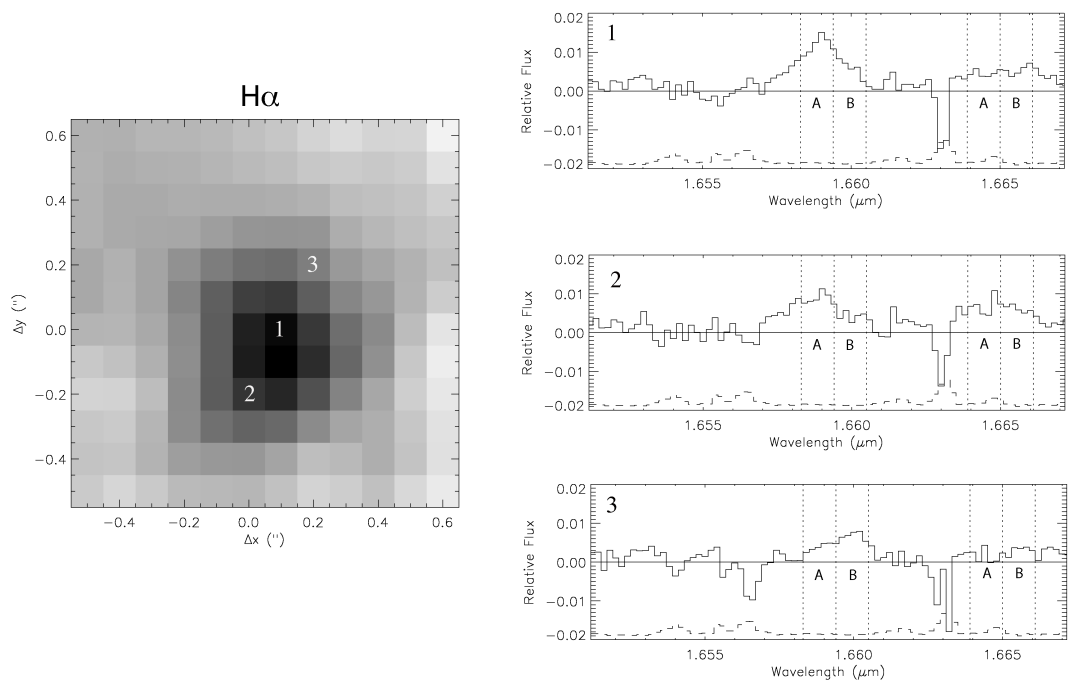}
\caption{\linespread{1}\small\textit{ 
\ha~image of Q2343-BX344 with individual lenslet spectra plotted (F$_{\lambda}$) for three different regions of the galaxy labeled 1, 2, and 3. The dashed line represents the standard deviation per spectral channel per lenslet off source. The region labeled `A' represents the \nii/\ha~ratio of a velocity location of \ha~having 0 \kms, and the `B'  region represents the \nii/\ha~ratio for a \ha~velocity location of 240 \kms. We present 2D \nii/\ha~ratio maps for each of these identified (A and B) spectral locations in Figure \ref{metalmaps}. 
}}\label{bx344_indiv_fig}
\end{center}
\end{figure*}

For HDF-BMZ1299 and Q2343-BX344, two dimensional velocity and dispersion maps were generated from their detected \nii~emission with the same Gaussian fit method as described for the \ha~emisison; see Figures \ref{bmz1299_fig} and \ref{bx344_fig} respectively. By matching spatial locations and kinematics of the \nii~gas to that of \ha~we find that the  \nii~emission is confined to a small region with very high ratios of \nii/\ha. We present individual spectra for three lenslet positions for HDF-BMZ1299 and Q2343-BX344 in Figure \ref{bmz1299_indiv_fig} and \ref{bx344_indiv_fig} to illustrate spatial locations where \ha~peaks, \nii~peaks, and on a lower S/N region of the galaxies. In Figure \ref{metalmaps}, we present \nii/\ha~maps for HDF-BMZ1299 and Q2343-BX344 at two different velocity locations with a spectrally collapsed width of $\Delta$$\lambda$=0.001 $\micron$ that are identified in their spectral locations in Figure \ref{bmz1299_indiv_fig} and \ref{bx344_indiv_fig}. Peak \nii~emission is offset from the peak \ha~emission spatially, where HDF-BMZ1299 has an offset of 0\farcs1 (0.86 kpc) and Q2343-BX344 has an offset of 0\farcs22 (1.9 kpc). For both of these sources the ratio of \nii/\ha~is highest at a redshift velocity associated with the peak location of \ha~emission. \nii/\ha~ratios close to $\sim$1 in a spatially concentrated region implies that both these galaxies have weak AGN activity, and are further discussed in \S\ref{indiv} and \S\ref{discuss}.

\subsection{Modeling Dynamics}\label{models}

Four of the objects (HDF-BMZ1299, HDF-BX1315, Q2343-BX344, and Q2343-BM145) exhibit a velocity gradient similar to a symmetrically rotating disk. For these galaxies we have modeled inclined disks and performed a least-squares fitting routine to their observed \ha~kinematics. Within this section we discuss our inclined disk model combined with beam smearing, and the caveats associated with these models of high-redshift disk candidates. We further investigate the effects of beam smearing with modeled intrinsic flux distributions of both a merger and an exponential disk. 

\subsubsection{Inclined Disk Modeling with Beam Smearing}\label{diskmodel}

The inclined disk model is built upon the routine developed within our previous paper on the disk candidate Q2343-BM133  \citep{wright07}, which uses a tilted ring algorithm for a symmetrically rotating disk \citep{bege87}. The disk model contains seven parameters, which are the center of rotation (x$_{o}$,y$_{o}$), position angle of the major axis ($\phi$), inclination angle (\textit{i}), velocity slope (\textit{m$_{v}$}), radius at which the plateau velocity is achieved (r$_{p}$) , and systemic velocity offset (V$_{o}$). At a given radius (R) from the center of the disk (x$_{o}$,y$_{o}$), a linear velocity profile is assumed with a rising circular velocity slope, until it reaches the plateau velocity (v$_{p}$) at a particular plateau radius (r$_{p}$), as shown by the following equation,

\begin{equation}
V_{c} = m_{v}R, ~~~~~ \mbox{for R \textless~r$_{p}$} \\ 
\end{equation}
 
\begin{equation}
V_{c} = m_{v}r_{p}, ~~~~~ \mbox{for R $\gtrsim$~r$_{p}$} \\
\end{equation}
~

Observed velocities at each resolution element are not independent of each other, and are blurred by the point spread function (PSF) of the LGS AO observations. To compare disk models to observed kinematics, we convolve each disk model with both the smoothing Gaussian kernel used on the observed data (FWHM = 0\farcs2) and an estimated point spread function (PSF) determined during each individual observation. We model the adaptive optics PSF by two Gaussians; one with the underlying seeing halo and one with a diffraction-limited core with a fraction of the flux equal to the Strehl ratio. Table \ref{obs} lists the estimated Strehl ratio used for each galaxy, which was calculated with the Keck AO performance estimator tool based upon the brightness of the tip-tilt star and separation from the science target. This final kernel (K$_{PSF}$) is produced by the convolution ($\ast$) of the smoothing Gaussian with the sum of the seeing halo (FWHM=0\farcs5) and core (FWHM=0\farcs1) Gaussians, and is represented by the following equation.

\begin{equation}
K_{PSF} = G_{smooth} * (G_{seeing} + G_{core}) 
\end{equation}
~

Extensive work has been invested to overcome many of the challenges with fitting disk models to local spiral galaxies (recent studies include \citealt{kraj06,jozsa07}). In particular, there is a known degeneracy between rotational velocity and inclination of the system as the product of v$_{c}$sin(\textit{i}). We find that disentangling the ambiguity between the two-dimensional model velocity field and the inclination is difficult. However, other disk parameters ($\phi$,\textit{m$_{v}$},r$_{p}$,V$_{o}$) can be well-determined, and in the case of high-redshift galaxies can help discern between a true disk and other systems (eg, merging, irregular galaxies). For each galaxy, we find the best fit for a disk model using an inclination of 45$^{\circ}$, which is the approximate average of available inclinations for these galaxies. We first perform a coarse grid search over the other six parameters with a set inclination of 45$^{\circ}$. Reduced-chi squared ($\tilde{\chi}^{2}$) were calculated for comparing each disk model solely with the observed velocities, where the degrees of freedom are calculated from the number of velocity measurements ($\sim$ 40 - 80) minus the number of parameters. At the lowest $\tilde{\chi}^{2}$, we perform a fine grid search around the six parameters to find the minimum $\tilde{\chi}^{2}$ for each galaxy and best fit disk model.

Another precaution is that observed velocity profiles are not only influenced by beam smearing, but also by the intrinsic flux distribution and inclination of the galaxy. This means that disk models are potentially biased by higher S/N regions within the system, which in turn may bias the centroid of the velocity profile. Within local spirals, typically mean surface brightness fades at larger radii. Although we have no prior knowledge of \ha~emission locations within these systems, we use higher S/N regions in each galaxy for our centroid values (x,y). We have found through our exhaustive grid search that derived x,y centers fall within one resolution element from the peak \ha~flux distribution.

\subsubsection{Results of Inclined Disk Models}\label{model_results}

\begin{figure*}[t]
\epsscale{0.8}
\begin{center}
\plotone{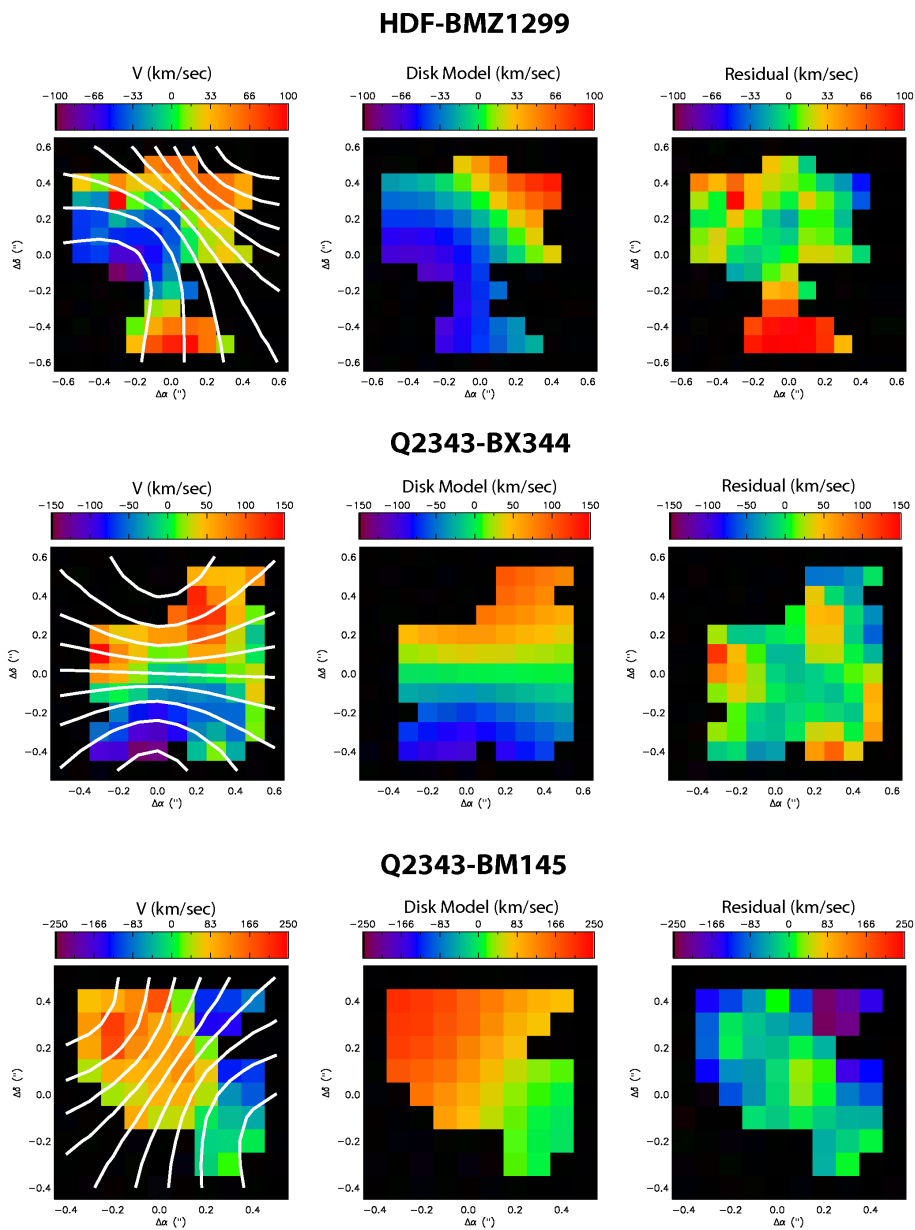}
\caption{\linespread{1}\small\textit{ 
Disk models and kinematics for the three disk candidates HDF-BMZ1299, Q2343-BX344, and Q2343-BM145. (Left) Observed kinematics with fitted disk model overlaid with a spider diagram. (Middle) Two dimensional disk model with the same spatial locations as observed kinematics. (Right) Residual map with the observed kinematics minus fitted disk model at each spatial location. 
  }}\label{diskfigs}
\end{center}
\end{figure*}

There are three galaxies (HDF-BMZ1299, Q2343-BX344, and Q2343-BM145) well fit by a disk model, and the fitted disk parameters are listed in Table \ref{diskmodel}. Subtracting the final disk model from observed kinematics yields a velocity residual map. An average residual is determined from the absolute value of the 2-D residual maps, and are 20 \kms, 27 \kms, and 46 \kms~ for HDF-BMZ1299, Q2343-BX344, and Q2343-BM145, respectively. In Figure \ref{diskfigs}, the three disk candidates are presented with their observed dynamics overlaid with their fitted spider diagram\footnotemark, model 2-D velocity map, and residuals from the disk model. The residuals for the disk model are comparable to the measured velocity uncertainties for each of these galaxies.  For each galaxy we determine the dispersion associated with their best fit disk model convolved with the observed PSF. We find that the disk model with beam smearing produces observed dispersions that peak at 66 \kms, 76 \kms, and 75 \kms for HDF-BMZ1299, Q2343-BX344, and Q2343-BM145. Observed dispersions of HDF-BMZ1299 and Q2343-BM145 are of the same order of disk-modeled beam-smeared dispersions. In contrast, Q2343-BX344 has observed dispersion that are three times greater than the modeled beam smeared dispersions, and implies this source likely has an intrinsic dispersion $\textgreater$ 100 \kms. 

\footnotetext{A ``spider" diagram refers to a velocity map classically representing rotation of a spiral galaxy.}

We also performed the disk model fit to HDF-BX1315 which shows a prominent, steep velocity gradient across the galaxy. This source has two resolved knots within the \ha~emission, and the velocity profile transitions from $\sim$+200 \kms~to -100 \kms~within one spaxel\footnotemark (0.86 kpc). In addition, the dispersion peaks between the two knots, as one might expect for two close in-falling sources with beam smearing. The  best disk model for this galaxy requires a very high velocity slope of m$_{v}$ = 379 km s$^{-1}$ spaxel$^{-1}$. The plateau velocity is achieved at a plateau radius of only 0\farcs09 (0.77 kpc) with a factor of three higher $\tilde{\chi}^{2}$ than our other disk candidates. Without the high spatial resolution of the LGS-AO system it would be difficult to discern from the kinematics that this is a merger system. Seeing-limited observations would likely lead one to interpret this galaxy as a rotating disk with a smooth velocity gradient and dispersion greatest at the center of the distribution.

\footnotetext{A ``spaxel" refers to the spatial resolution element on-sky from an integral field spectrograph. For OSIRIS, a spaxel refers to the spatial scale for the lenslet array.}

\subsubsection{Model Flux Distributions: Two Component Merger and Exponential Disk}\label{fluxmodel}

To further investigate beam smearing and to discriminate between disk and merger systems, we model the \ha~flux distribution for both a two component merging system and an exponential disk. We have selected velocity profiles and flux orientations for two of our sources, HDF-BX1315 and Q2343-BX344. This allows an exploration of modeled flux distributions while offering direct comparisons to our observed dynamics, which are then fully generalized to the rest of the sample. 

\begin{figure*}[t]
\epsscale{0.8}
\begin{center}
\plotone{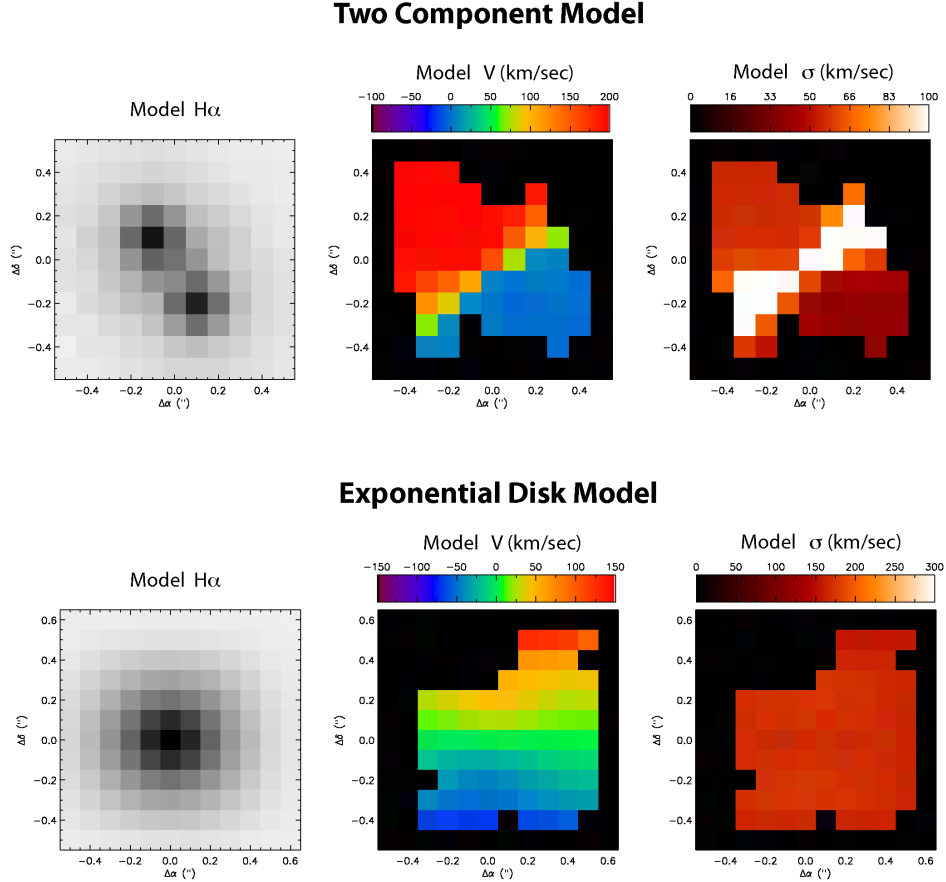}
\caption{\linespread{1}\small\textit{ 
To investigate beam smearing further we have generated model flux distributions of a two component merger system and an exponential disk model and have combined them with a velocity profile, and then convolved them with our estimated PSF. We have used HDF-BX1315 and Q2343-BX344 orientations (PA) on sky and their velocity profiles to model their 3D flux distribution. TOP PANEL:  (Left) Two single resolution (0\farcs1) knots were generated in a 3D cube, at the peak locations we observe in HDF-BX1315 with a similarly observed velocity offset and intrinsic dispersion of 45 \kms.  The model two-component data cube was then convolved with our estimated PSF (Strehl = 30\%) and smoothing profile (FWHM=0\farcs2). We present the blurred \ha~flux distribution of a two knot system, beam smeared velocity offset and dispersion maps. Note that the velocity gradient and dispersions match closely to what is observed in HDF-BX1315 in Figure \ref{acs}. BOTTOM PANEL: An exponential flux distribution with an inclination of 45$^{\circ}$ was generated in a 3D cube with a similar velocity offset and intrinsic dispersion of 150 \kms. We present the blurred \ha~flux distribution of an exponential disk, beam smeared velocity offset and dispersion maps. }}\label{modelfig}
\end{center}
\end{figure*}

For the two component system, we generate a flux distribution of two single resolution-element (0\farcs1) knots at the locations of the two \ha~peak locations within HDF-BX1315 with a similar flux ratio and distance. A model OSIRIS cube was generated for these two knots with a velocity offset and intrinsic dispersion of $\sigma$$_{\rm{int}}$$\sim$50 \kms similar to that of HDF-BX1315, with the instrumental width added in quadrature. Using the estimated AO PSF (Strehl=30\%) for HDF-BX1315, we convolve each wavelength slice of our two component cube with the PSF and smoothing Gaussian (FWHM=0\farcs2). We present the beam smeared model \ha~flux distribution, velocity offsets, and velocity dispersions (instrumental width removed) in Figure \ref{modelfig}. The steep velocity gradient seen across the two component model matches well with the observed velocity structure of  HDF-BX1315, and with the dispersion peak around $\sim$100 \kms~between the two knots as seen in our source. This further implicates HDF-BX1315 as a merging system with apparent kinematics being explained by beam smearing.  

For the exponential disk, we chose to model the characteristics of Q2343-BX344 with its best fit disk position angle, inclination of 45 degrees, and velocity profile. We used a similar procedure to that described for the two component model. We generated a modeled OSIRIS cube of an exponential disk with Q2343-BX344 disk velocity profile, with an assumed intrinsic velocity dispersion of $\sigma$$_{int}$=100 \kms and scale length of 0.9 kpc. This modeled cube was convolved with the estimated AO PSF (Strehl=23\%) and smoothing Gaussian (FWHM=0\farcs2), and its \ha~flux distribution and kinematics are presented in Figure \ref{modelfig}. The velocity profile of an exponential disk flux distribution with beam smearing lowers the observed plateau velocities (v$_{p}$) of the system, since the central velocities with bright flux are blended with the outer wings of the galaxy. Beam smearing hence reduces the observed plateau velocities, and the effect is particularly strong if the PSF is comparable in size to the object. In our model system, the final beam smeared plateau velocity is approximately 30\% lower than the assumed physical velocity due to this effect. 

Following the prescription above for modeling the flux distribution of exponential disks, we find a similar amount of softening has occurred in the other disk candidates, and therefore the inclined-disk modeled plateau velocities corrected for beam smearing increase for HDF-BMZ1299, Q2343-BX344, and Q2343-BM145 to v$_{p}$=167 \kms, v$_{p}$=198 \kms, and v$_{p}$=195 \kms. In contrast, beam smearing is unable to explain the large dispersions observed within Q2343-BX344. The modeled dispersions of an exponential disk with beam smearing instead show a relatively flat dispersion of $\sim$165$\pm$10 \kms~across the whole source. In order to match our observed dispersions of Q2343-BX344, we need to assume intrinsic dispersion of the system $\sigma_{\rm{int}}$=150 \kms. This matches our conclusion above that a large fraction of the observed dispersions of Q2343-BX344 must be intrinsic to the source. However, for the other two disk candidates, HDF-BMZ1299 and Q2343-BM145, we can easily generate their observed dispersions with low intrinsic dispersion $\sigma_{\rm{int}}$$\lesssim$10 \kms, which produces observed dispersions of order $\sigma_{\rm{obs}}$$\sim$70 to 100 \kms. Given the simplified disk model this is in excellent agreement, and implies these two systems may be sustaining star formation within a ``relaxed" disk system. 

There are several caveats with the modeled flux distributions and beam smearing models that should be highlighted. One of the largest uncertainties is whether the flux center corresponds to the dynamical center of the disk system. If the flux center does not match the dynamical center then beam smearing plays a stronger role, since the dynamics within the brighter regions will smooth and confuse the kinematics at lower S/N regions. As discussed above, another consideration is the inclinations of these systems. For instance, an edge-on disk system with beam smearing will produce larger observed dispersions than a face-on disk system, so coupling the observed dispersions with disk modeling may be a potential way of disentangling the degeneracy with inclination. More intricate flux distributions with varying intrinsic dynamics and the effects of beam smearing will be investigated in a future paper (Wright et al, in prep). We further discuss implications of high intrinsic dispersions and derived plateau velocities of our disk systems in \S\ref{discuss}.
 
\subsection{Mass Estimates}\label{mass}

We use observed \ha~flux emission and dynamics to determine the masses of both the merger and disk candidate galaxies using several different methods. There are a number of assumptions used for these calculations, which are likely to have high uncertainties of a factor of 2 to 4 in their estimated values. We attempt these mass estimates to make suitable comparison between our z$\sim$1.6 sources and other high-redshift galaxy samples.  

\subsubsection{Virial Mass Estimates}\label{virialmass}

There are several methods for investigating mass estimates of these galaxies using observed kinematics of the velocity profile and dispersion. Most studies have not spatially resolved the internal motions, and have used the global velocity dispersion in conjunction with the virial theorem to determine a mass estimate. This requires numerous assumptions regarding the galactic system, primarily that the system is bound and dynamically relaxed in a particular geometry and velocity profile (eg, isothermal sphere, disk, elliptical). Using the observed velocity dispersion, authors have often used the virial theorem to calculate total mass of the system with the following equation,

\begin{equation}
M_{virial} = \frac{C\sigma_{v}^{2}r_{1/2}}{G}.
\end{equation}

The constant factor (C) is determined from integrating the mass distribution and velocity profile of the system within the half-light radius (r$_{1/2}$) . Previous authors have assumed varying constant factors representing different physical orientations. A simple geometry which has a spherically uniform density profile has a constant factor C=5 and has been used previously for virial mass estimates \citep{pett00,erb03,erb06a}. To represent a uniform density disk with an  inclination of 45$^{\circ}$, a constant factor of C=3.4 has been used \citep{erb06a, wright07, law07}. However, virial masses for disk galaxies are more complicated to estimate since the calculation involves integrating an exponential mass distribution of a disk with a non-isotropic velocity profile. Therefore, for purposes of comparison between merger and disk candidates we have chosen to use the constant C=3, which assumes a spherical relaxed system where luminosity traces the mass with a de Vaucouleurs r$^{1/4}$ profile within an effective half-light radius (r$_{1/2}$). These virial mass estimates are similar to mass estimates for local elliptical galaxies. This simplified estimate would not hold true if these systems were in higher density environments or had extensive tidal stripping and merging. But given our limited knowledge of how the luminosity maps to the underlying mass, our measured dispersion and half-light radius with a de Vaucouleur's profile offers a suitable comparison between our sample and other studies at similar epochs.

The half-light radius (r$_{1/2}$) was determined for each resolved knot using a growth curve generated from aperture photometry, where the center of the object was assumed to be at the \ha~flux peak. Table \ref{kinemass} includes the half-light radius determined for each source and the calculated enclosed virial mass. Virial mass estimates span an order of magnitude across our sample, from the smallest virial mass for the subcomponent HDF-BX1315-NE with 1.5x10$^{9}$ \msun, to the highest of 8.5x10$^{10}$ \msun for Q2343-BX344. Our mass estimates are typically M$_{\rm{vir}}$$\sim$1-2x10$^{10}$ \msun, and are somewhat comparable to \citealt{schreib06} (0.1-10x10$^{10}$ \msun) seeing-limited sample, even though their r$_{1/2}$ are much larger in size. More appropriately comparing our study to another LGS-AO study for z$\sim$2-3 galaxies, \citet{law07}, we find our galaxies have larger half-light radii with similar observed dispersions, and hence we find higher virial masses. 

\subsubsection{Enclosed Mass Estimates}\label{enclosedmass}

Using the derived disk models for our disk candidates, we are able to directly calculate the enclosed masses of these systems. Assuming Keplerian motion of a highly flattened spheroid, we calculate the enclosed dynamical mass for HDF-BMZ1299, Q2343-BX344, and Q2343-BM145 with the following equation,

\begin{equation}
M_{enclosed} = \frac{2 v_{c}^{2}r}{\pi G} 
\end{equation}

We use the plateau velocity (v$_{p}$) from the disk models corrected for beam smearing as the circular velocity (v$_{c}$), and for the radius (r) we select the farthest distance from the dynamical center which has observed \ha~kinematics.  We find an enclosed mass of 2.6x10$^{10}$,  3.9x10$^{10}$, and 3.1x10$^{10}$ \msun for HDF-BMZ1299, Q2343-BX344, and Q2343-BM145, respectively. For two of the disk candidates, HDF-BMZ1299 and Q2343-BM145, enclosed and virial mass estimates agree well with each other, whereas Q2343-BX344 has a virial mass which is $\sim$2 times larger than the enclosed mass. Virial mass calculations for either disk or merger candidates are the least robust estimates since numerous assumptions are made for the system and hence have large uncertainties. 

\subsubsection{Halo Mass Estimates}\label{halomass}

Numerous dark matter simulations have correlated the distribution of clustering of galaxies to that of clustering of dark matter halos, and have related the circular velocity of dark matter to that of the total halo mass. Our calculated enclosed masses are lower limits to total masses of the disk candidates, since they underestimate the halo mass of the galaxy which includes stellar, gas, and dark matter mass components beyond the assumed radius (\textit{r}). In addition, since these galaxies are gas rich they likely have large reservoirs of cool gas beyond the radius where we have observed \ha~flux. We estimate the halo mass (M$_{\textrm{halo}}$) using a spherical virialized collapse model \citep{peebles80, white91,mo02} with the concordance cosmology, and the assumption that observed circular velocities trace the modeled halo circular velocities.
 
\begin{equation}
M_{halo} =  0.1 H_{o}^{-1} G^{-1} \Omega_{m}^{-0.5} v_{c}^{3} (1+z)^{-1.5} ~\mathrm{(M_{\sun})}
\end{equation}

For our three disk candidates, using plateau velocities from their disk models with beam smearing corrected, we find total halo masses of 0.7$\times$10$^{12}$ \msun, 1.2$\times$10$^{12}$ \msun, and 1.1$\times$10$^{12}$ \msun for HDF-BMZ1299, Q2343-BX344, and Q2343-BM145. All of the halo masses are close to the observed Milky Way's total halo mass \citep{wilk99} . Compared to the latest results from \citet{kom08} with the fraction of baryonic matter to total matter $\Omega$$_{b}$/$\Omega$$_{m}$=0.171, we find that our sources have M$_{encl}$/M$_{halo}$ less than 0.05. This ratio assumes that our enclosed mass calculations are primarily dominated by baryons, and follows a similar estimate performed for the higher redshift sources of \citet{schreib06}. If we assume that these galaxies have assembled the majority of their baryonic mass within their dark matter halo at z$\sim$1.6, then these galaxies have a fraction of baryonic mass to total mass slightly less to that of the present-day Milky Way, of 0.08$\pm$0.01 \citep{cardone05}. Our z$\sim$1.6 galaxies are also in agreement with the halo masses found in \citet{adel05}, which used the dependence of halo clustering strength on mass to find a characteristic halo mass, for the large binned galaxy distribution of z$\sim$1.7$\pm$0.4. Although it is interesting to compare our inferred halo masses to other studies, it is important to note that these halo mass estimates have significant uncertainties and are strongly coupled with the latest dark matter and clustering models.

\subsubsection{Gas Mass Estimates}\label{gasmass}

Present-day star forming galaxies are known empirically to follow a Kennicutt-Schmidt (K-S, \citet{kenn98}) law where the surface density of gas to star formation follows a power law above some critical gas density ($\Sigma_{SFR}$ = A$\Sigma_{gas}$$^{n}$). The K-S law has been well studied and modeled for nearby galaxies, but has only recently been investigated at higher redshift within the \citet{erb06b} z$\sim$2 sample. They were able to combine their measured SFR with the radius of \ha~from near-infrared long-slit spectroscopy to infer gas surface density and hence a total gas mass for each of their objects. 

Two-dimensional \ha~flux distributions allow us to directly estimate the gas surface density ($\Sigma_{gas}$) and total gas masses (M$_{gas}$) for all of our sources. For disk and merger candidates, we assume an inclination of 45$^{\circ}$ to map the 2D surface gas density. A dereddened \ha~luminosity is found for each spatial element (i, j) using the SED fit E(B-V) measurement, and is then converted to luminosity per unit area:

\begin{equation}
L_{H_{\alpha_{i,j}}} = \frac{4 \pi D_{L}^{2}~cos(i)~F_{H_{\alpha_{i,j}}}} {k^{2}} ~~~\mathrm{(ergs~s^{-1}~kpc^{-1})}
\end{equation}

\noindent{Luminosity distance (D$_{L}$), dereddened \ha~flux at each spatial element (F$_{H_{\alpha_{(i,j)}}}$), scale conversion (k) of kiloparsecs to arcseconds, and the inclination angle (i) are used. We then use the KS relation between luminosity per unit area to surface gas density for each spatial element.}

\begin{equation}
\Sigma_{gas_{i,j}} = 1.6\times10^{-21} (L_{H_{\alpha_{i,j}}})^{0.71} ~~~\mathrm{(M_{\sun}~kpc^{-1})}
\end{equation}

\noindent{We are able to then estimate total gas mass (M$_{gas}$) by summing each spatial element of the surface gas density at its projected inclination, with the scaling conversion (k) and inclination angle (i). }

\begin{equation}
M_{gas} = \sum_{i,j} \frac{k^{2}~\Sigma_{gas_{i,j}}}{cos (i)} ~~~\mathrm{(M_{\sun})}
\end{equation}

A range of total gas masses are found between 1-4.4$\times$10$^{10}$ \msun with large statistical errors (see Table \ref{kinemass}). Using the stellar mass found from SED fits, we find the fraction of gas compared to the total baryonic matter by using $\mu$ = M$_{gas}$ / (M$_{gas}$ + M$_{stellar}$), and find an average gas fraction of 78\% for all our galaxies. Our galaxies are gas rich and have not yet converted the majority of their baryonic mass into a stellar population. Although we have only six sources, we find a similar trend as seen in \citet{erb06a}, where lower gas fractions correspond to higher stellar masses. For our three disk candidates we can compare the dynamically enclosed mass (M$_{enclosed}$) to the total baryonic matter (M$_{gas}$ + M$_{stellar}$), and all fall within the 4$\sigma$ level on \citet{erb06a} z$\sim$2 relationships of gas mass to stellar mass parameters. It should be noted there are a number of assumptions required to use the K-S law to calculate M$_{gas}$. In particular, we assume an inclination of 45$^{\circ}$, and a global extinction across the entire galaxy inferred from the SED fits. Given the large uncertainties for gas and stellar mass estimates, these gas fractions are uncertain. However, we believe our gas fractions to be lower limits, since SED fits come from seeing-limited photometry that encompasses a larger region than our \ha~measurements and since there are likely regions of gas that are not actively forming stars at a sufficient rate for us to detect, which further indicates these systems are highly gas rich.

\section{UV Interstellar and Emission Kinematics}\label{uv}

\begin{figure*}[t]
\epsscale{0.8}
\begin{center}
\plotone{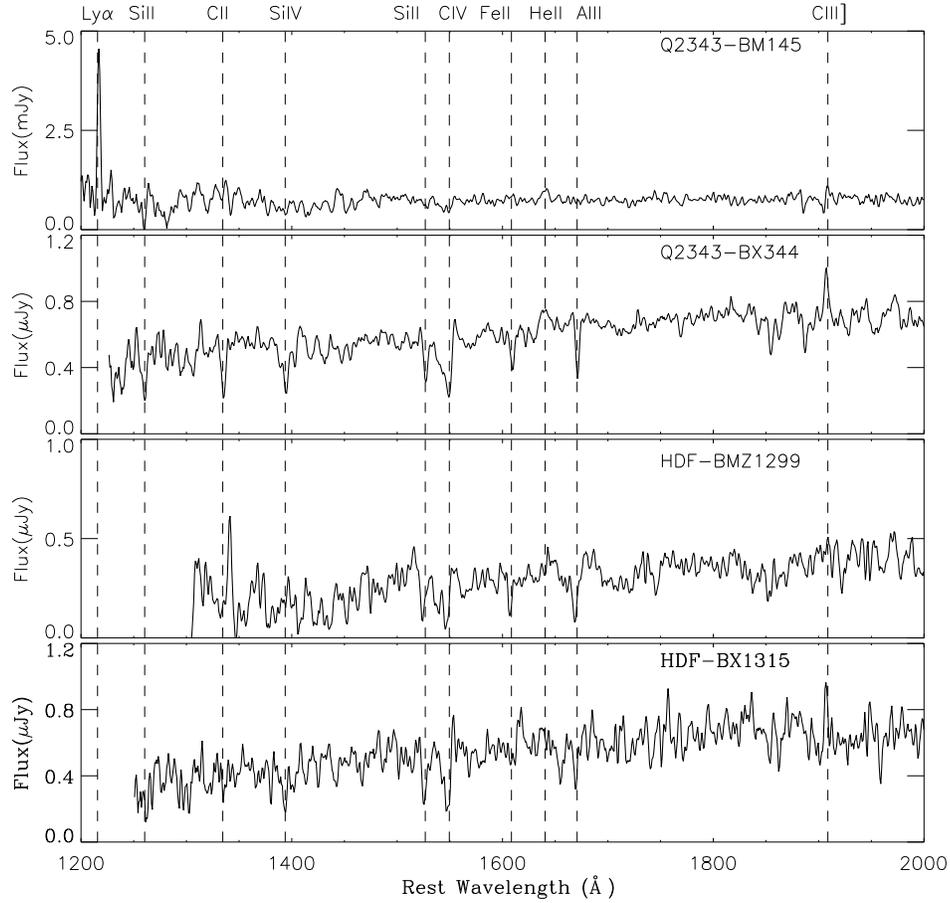}
\caption{\linespread{1}\small\textit{ 
Flux calibrated UV spectra of four galaxies: Q2343-BM145, Q2343-BX344, HDF-BMZ1299, and HDF-BX1315. Each spectrum has been smoothed by a boxcar 3 filter and shifted to rest-frame wavelengths based on
their observed \ha~redshift. Vertical dotted lines indicate locations of some of the most prominent interstellar absorption and emission lines. 
  }}\label{uvspectra}
\end{center}
\end{figure*}

For our redshift regime at z$\sim$1.6, rest-frame UV wavelengths between $\sim$1200\AA~to 2200\AA~are redshifted into the prime wavelength interval for optical spectroscopy. The six galaxies herein were observed with the LRIS-B spectrograph over a four year time interval. An instrumental setup similar to that described in \citet{steidel04} was used with the 400 groove mm$^{-1}$ grism blazed at 3400\AA~spanning the observed wavelengths of 3100\AA~to 6500\AA. Spectra were taken with varying integration times, resulting in high signal-to-noise spectra for HDF-BX1315, HDF-BMZ1299, Q2343-BX344, and Q2343-BM145 presented in Figure \ref{uvspectra}. Our spectra resembled those of \citet{steidel04} for objects at a similar redshift regime, which showed varying strengths of interstellar absorption lines (eg, Si{\sc ii}, O{\sc i}, C{\sc ii}, S{\sc iv},C{\sc iv}, Al{\sc iii}, and Fe{\sc ii}) and emission features (eg, Ly$\alpha$, He{\sc ii},Cl{\sc ii}). For each line with sufficient signal-to-noise in each galaxy, we measured redshift and velocity offset with respect to the observed \ha. Mean velocity offset for interstellar absorption and emission lines for each galaxy are presented in Table \ref{lris}. Ly$\alpha$ emission was detected only for Q2343-BM145, with a redshifted velocity offset of v$_{Ly\alpha}$=316 \kms~ with respect to its nebular \ha~emission. Three of the galaxies have typical blueshifted interstellar absorption lines of $\sim$-260 \kms, which is postulated to be from large scale outflows from star formation activity \citep{erb03,steidel04}. In contrast, Q2343-BX344 is atypical with the opposite velocity structure, with redshifted interstellar absorption lines of 144 \kms~and blueshifted emission lines of -122 \kms which may represent in-falling gas onto this galaxy. Kinematics of rest-frame UV interstellar lines do not trace the dynamics of \ha~emission which is dominated by \hii~regions and star formation, but rather traces the inflow and the more typically seen outflows of gas that are generated by ``superwinds" from star formation activity \citep{steidel04}. As we have shown in \S\ref{sfr}, this picture is consistent given our SFR per square-kiloparsec for all of our sources, would have significant winds across the entire observed \ha~regions of the galaxy. 

\section{Individual Galaxies}\label{indiv}

In this section we summarize the results for each galaxy, and expand on the discussion of their metallicity, mass distributions, kinematics, formation mechanisms, and likely evolution to present-day sources. Although we only have a small number of sources, there is an extraordinary amount of information from combining \ha~and \nii~dynamics and morphologies with UV spectroscopy and broadband photometry, which merits discussing each galaxy separately.

\subsection{HDF-BX1287}\label{bx1287}

There are two distinct components within the \ha~flux distribution of HDF-BX1287, with a large bright southern component and a secondary source north-west by 1\farcs46 (12.5 kpc), which likely represent an active merger system with the largest separation in our sample. HDF-BX1287 images and kinematics are presented in Figure \ref{acs}.  Within the ACS rest-frame UV images, the northern source is clearly resolved and the southern component is two separate resolved components, with fainter emission extending from both sides of the southern components. The \ha~flux distribution primarily spans $\sim$1\farcs1 (9.4 kpc) across the southern source with a slight velocity gradient of $\sim$100 \kms, and faint \ha~emission off the south-east and south-west edges of the source. The northern component has a flat \ha~flux distribution giving rise to a large half-light radius (r$_{1/2}$) of 3.9 kpc,  with a velocity offset from the southern component of v$\sim$200 \kms.  There is very little velocity gradient of the northern source, with a low dispersion of $\sigma$=45 \kms. The peak dispersion spatially matches the peak of the flux distribution, which may be indicative of a minor merger within the northern source.

Individual virial masses for northern and southern sources of HDF-BX1287 are M$_{vir}$ = 0.6x10$^{10}$ \msun and M$_{vir}$=1.4x10$^{10}$ \msun. The total stellar mass M$_{*}$=2x10$^{9}$ \msun for HDF-BX1287 was inferred from integrated light from both the northern and southern sources. This object has the highest gas fraction of $\mu$=95\% for all of our sources, with a total gas mass of M$_{gas}$=6.1x10$^{10}$ \msun. Using the large distances between the northern and southern components and assuming Keplerian motion, we calculate a total mass of 4.2x10$^{10}$ \msun. Total inferred mass of the system is therefore 30\% less than the measured gas mass, and even given the large uncertainties with gas mass estimates this likely reflects only a minimum dynamical mass. We find a merging time scale of 1.1 Gyr with t$_{m}$ = (2$\pi$r/v$_c$)(M/m)  \citep{graham90}, where we assume that the mass ratio (M/m) of the system is the ratio of their individual virial mass estimates. The most striking mass comparison is between the virial mass to the total mass estimate, since the virial mass underestimates the total mass of the system by at least a factor of 2, but falls within the uncertainties for all these mass estimates.
 
\subsection{HDF-BX1315}\label{bx1315}

OSIRIS observations of HDF-BX1315 highlight the high spatial resolution necessary for unraveling intrinsic dynamics of high-redshift galaxies. \ha~dynamics of this galaxy show a velocity gradient of 300 \kms~with a steep velocity transition within one spatial element ($\lesssim$0.86 kpc). \ha~emission shows two resolved sources in the north-east (NE) and south-west (SW) of the flux distribution. Rest-frame UV imaging shows three distinct sources that are each elongated with a cigar-like shape. The northern-most source was unfortunately on the edge of our OSIRIS dither pattern and was therefore not detected. Figure \ref{acs} illustrates the location of the \ha~observations on the ACS image and presents the \ha~kinematics. As previously described in \S\ref{model_results}, we attempted to fit inclined disk models which had the highest $\tilde{\chi}^{2}$ with a large, unphysical velocity slope m$_{v}$$\gtrsim$ 350 \kms~within less than one resolution element (0.51 kpc). There is a flat velocity gradient on each of the resolved NE and SW knots, with the highest dispersion of $\sigma$$\sim$220 \kms~peaking between the two knots. Dispersion of each resolved knot is relatively low, with $\sigma$=32 \kms~for the NE and $\sigma$=65 \kms~for the SW components. Assuming that the mass ratio is the ratio of each components virial mass, we follow the same calculation for HDF-BX1287, and find a merging time scale for these two components of $\sim$660 Myr. This source has the lowest gas fraction (60\%) of any of our sources, but this is likely underestimated since our \ha~observations missed the third component seen in the ACS image. Based on HDF-BX1315 \ha~and rest-frame UV flux distribution and kinematics, we believe this galaxy is a merging system, with a large stellar mass of M$_{*}$=1.5x10$^{10}$ at this epoch, and a large reservoir of gas yet remaining.

\subsection{HDF-BMZ1299}\label{bmz1299}

In Figure \ref{bmz1299_fig}, we present HDF-BMZ1299 OSIRIS observations of \ha~and \nii~emission and kinematics, with the HST ACS color-stacked image. Primary \ha~flux distribution is confined within 1\farcs0 (8.6 kpc), but shows fainter \ha~flux extended to the north and to the west, giving a full extent of the galaxy of 1\farcs4 (12.1 kpc). Within the HST images, there is a distinct, compact UV knot ($\sim$0\farcs4) north of the primary source which coincides with faint, unresolved \ha~emission from OSIRIS observations. \ha~kinematics show a significant velocity gradient of $\sim$150 \kms~at position angle 0$^{\circ}$. Kinematics near the UV knot are consistent with overall global motion of the galaxy, and does not have a higher velocity dispersion peak as one might expect if this was an independent object with its own dynamics. Therefore, this may represent an extreme mode of star formation occurring within the disk or a lane of higher extinction, and not a separate colliding galaxy. Unlike our merger candidates, the velocity dispersion is close to the minimum at the \ha~flux peak ($\sim$40 \kms) and remains relatively flat across the source.  This object had the lowest Strehl ratio achieved and therefore should have the highest effects of beam smearing, yet among all of our disk systems we see the lowest dispersions across this source. The western-most component of the HDF-BMZ1299 has the ``tail" in the ACS image, and has significant velocity deviations (v$\sim$95 \kms) from the global velocity gradient. This region was masked out when determining the disk model that best fits the observed dynamics (see Figure \ref{diskfigs}), and we find a plateau velocity of v$_{p}$=111 \kms~within 0\farcs09 (0.77 kpc), and corrected for beam smearing plateau velocities of v$_{p}$=167 \kms. We believe this source is a good candidate for a disk system, given our $\tilde{\chi}^{2}$ of 1.2, low residuals to our fit of order $\sim$20 \kms, and observed dispersions that are in reasonable agreement with expected dispersions for an inclined disk with our observed PSF.

\begin{figure*}[t]
\epsscale{0.8}
\begin{center}
\plotone{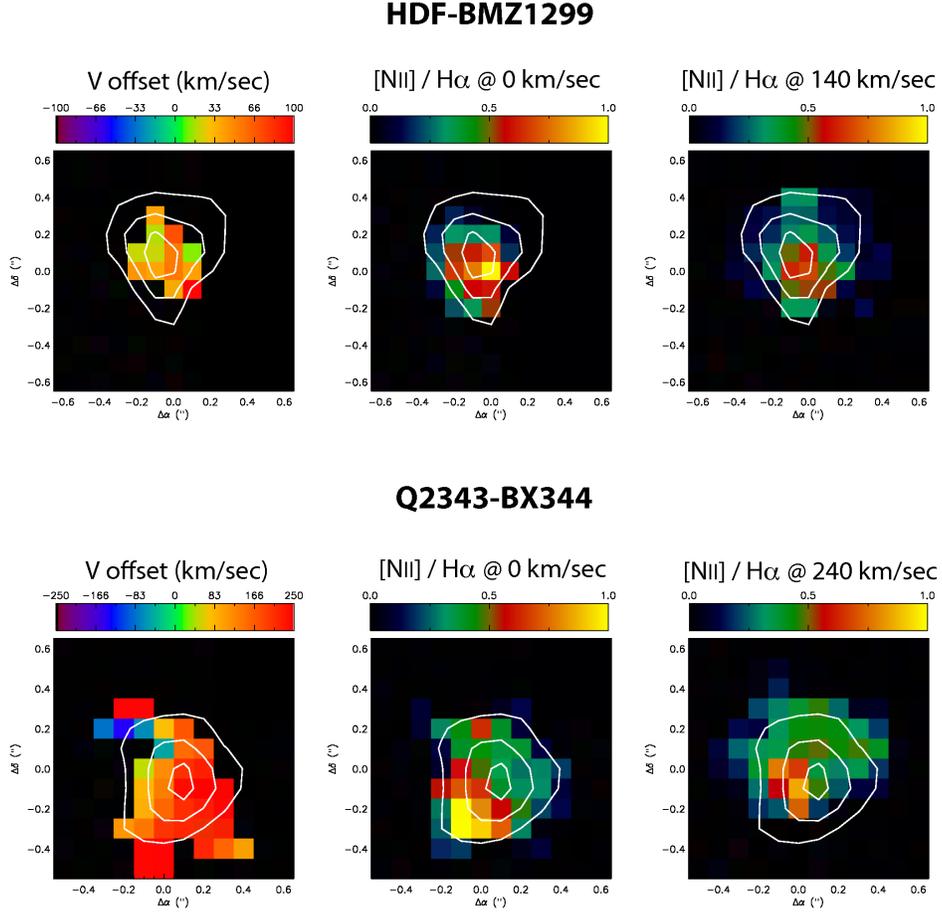}
\caption{\linespread{1}\small\textit{ 
\nii/\ha~ratio maps for HDF-BMZ1299 and Q2343-BX344 are presented to illustrate locations of high and low nebular ratios at different velocity locations. Contours of \ha~emission are superimposed on each image. (LEFT) \nii~velocity offsets with respect to the mean \ha~redshift for each source. HDF-BMZ1299 and Q2343-BX344 both have \nii~emission that is redshifted with respect to \ha~emission. (MIDDLE) Ratio of \nii/\ha~at the peak location of \ha~emission (0 \kms), with a collapsed spectral region of $\Delta$$\lambda$=0.001$\micron$. (RIGHT) Ratio of \nii/\ha~at the peak location \nii~emission with a collapsed spectral region of $\Delta$$\lambda$=0.001$\micron$. There is a  high \nii/\ha~ratio concentrated in one to two spatial elements for each source, with lower \nii/\ha~values occurring elsewhere within the galaxuy. This implies there is a central AGN within each of these sources, indicated by the locations of the high \nii/\ha~locations. }}\label{metalmaps}
\end{center}
\end{figure*}

HDF-BMZ1299 is one of two sources that has bright \nii~($\lambda$6585) emission, with total \nii~flux of 6.3x10$^{-17}$ ergs s$^{-1}$ cm$^{-2}$ contained within 0\farcs5 (4.2 kpc). In Figure \ref{bmz1299_fig}, \nii~flux distribution, velocity profile, and velocity dispersions are presented.  In Figure \ref{metalmaps}, \nii/\ha~ratio maps are presented at two different velocity locations within the source, with a map of the \nii~velocity offset with respect to the \ha~emission. Peak \nii~emission does not match the peak \ha~spatial location and is instead offset 0\farcs1 to the north-west, and the \nii~emission is also redshifted $\sim$50 \kms~from the expected location based on \ha~redshift. There is a slight \nii~velocity gradient from 10 to 100 \kms~running from north-east to south-west with dispersions as high as $\sigma$$\sim$150 \kms. In \S\ref{metal}, we find a high ratio of \nii/\ha~of 0.9 mapped to the same spatial locations, as reported in Table \ref{global}, which is well above the ratio for ``normal" star forming galaxies. However, if we used the global \nii~flux and the global \ha~flux of the system, as would be performed for slit spectroscopy, we would find an \nii/\ha~ratio closer to 0.3. This nebular ratio is closer to the \nii/\ha~ratios determined from the z$\sim$2 sample in \citealt{shap05}, and higher than their z$\sim$1.5 ratios. As seen in Figure \ref{metalmaps}, given that the high \nii/\ha~ratio of 0.9 is localized within a small spatial region (0\farcs1), and the high \nii~dispersions, we believe that the majority of the \nii~emission is generated by the presence of an AGN. Since the AGN is not the dominant emission mechanism seen in rest-frame UV spectroscopy and \ha, it is only found because of the high spatial resolution achieved with OSIRIS and the LGS-AO system. This suggests that prior metallicity estimates from global line ratios by slit spectroscopy may be biased by the presence of relatively weak AGN. In addition, the \nii~kinematics of slight velocity gradient and high dispersion might reflect the direction of AGN induced outflow in a westerly direction. 

\subsection{Q1623-BX491}\label{bx491}

\ha~emission arises in two distinct knots separated by 0\farcs6 (5.0 kpc) in the north-east (NE) and south-west (SW) portions of Q1623-BX491. \ha~flux distribution and kinematics are presented in Figure \ref{twoobj}. Velocity offsets of this galaxy are disordered and suggest a merger scenario. There is not a significant velocity gradient between the two knots, but irregular velocities as high as v$\sim$200 \kms~exist in the north-west of the galaxy. Since the velocity profile is relatively flat between the two resolved knots, it's likely that kinematics of these knots are primarily within the plane of the sky. Similar to our other merger candidates, the dispersion of $\sigma$$\sim$160 \kms~peaks between the two \ha~knots, and then rapidly falls to $\sigma$$\sim$ 70 \kms~ at each of the knots, where it likely represents intrinsic motion of the gas. Virial mass for Q1623-BX491 is 2.5 times less than the total baryonic matter inferred from the stellar mass and gas mass. There was not sufficient signal-to-noise in the LRIS spectroscopy to obtain useful rest-frame UV absorption and emission line characteristics. 

\subsection{Q2343-BX344}\label{bx344}

As seen in Figure \ref{bx344_fig}, \ha~emission for Q2343-BX344 extends over an arcsecond,  and there is a prominent velocity gradient at position angle of 0$^{\circ}$ from -150 to +150 \kms. Q2343-BX344 is well-fit by an inclined disk model with the lowest $\tilde{\chi}^{2}$ of all our disk candidates, and low residuals of $\sim$27 \kms~between the model and observed velocities. In Figure \ref{diskfigs}, we present the modeled disk, spider diagram overlaid to the observed kinematics, and 2D residuals. A plateau velocity v$_{p}$ of 132 \kms~is attained within 0\farcs13 (1.1 kpc), and plateau velocities corrected for beam smearing of v$_{p}$ of 198 \kms. The major axis of the galaxy spans 6.1 kpc, with a minor axis of 5.3 kpc. Q2343-BX344 shows high dispersions across the disk of $\sigma$$\sim$150 \kms, where the peak dispersion location ($\sim$260 \kms) is offset from the peak \ha~emission. These elevated dispersions far exceed the expected dispersion based upon a simple exponential disk model with beam smearing, and likely are intrinsic to the system and potentially reflect a ÒpuffyÓ or ÒheatedÓ disk system. 

Prominent \nii~($\lambda$6585) emission is detected in Q2343-BX344 covering $\sim$0\farcs9x0\farcs8 region, with the centroid displaced from peak \ha~by $\sim$0\farcs2 (1.7 kpc) towards the southeast. In the bottom panel of Figure \ref{bx344_fig}, \nii~flux distribution, velocity profile, and velocity dispersions are presented. In Figure \ref{metalmaps}, \nii/\ha~ratio maps are presented at two different velocity locations within the source, with a map of the \nii~velocity offset with respect to the \ha~emission. This galaxy's \nii~emission features resemble those of HDF-BMZ1299, with \nii~emission redshifted with respect to \ha~and \nii~high dispersions. We see dispersions as high as $\sigma$$\sim$300 \kms~off-center from the \nii~peak, which is closer spatially to the center on the \ha~peak emission. \nii~is kinematically distinct from \ha, with a velocity profile that runs perpendicular to the \ha~velocity profile with $\pm$100 \kms. Rest-frame UV emission lines are blueshifted with respect to \ha~by $\sim$-120\kms. Using corresponding spatial locations of \nii~and \ha, we find a high \nii/\ha~ratio of 0.8. If we instead use global fluxes of both \nii~and \ha, we find a lower ratio of 0.4 for \nii/\ha, similar to our results for HDF-BMZ1299. Looking at Figures \ref{bmz1299_indiv_fig} and \ref{bx344_indiv_fig}, there are two peaks for both \ha~and \nii~emission seen with the individual lenslet spectra. As seen in Figure \ref{metalmaps} the \nii/\ha~ratio is concentrated within a 0\farcs2 spatial region offset to the south-east of the peak \ha~emission. This suggests there is an AGN contribution within this region, and there is underlying star formation through the rest of the galaxy which dominates the overall luminosity.

\subsection{Q2343-BM145}\label{bm145}

Q2343-BM145 is a single compact source with its primary \ha~flux distribution being contained within 0\farcs5 (4.3 kpc), and faint \ha~flux extending towards the SW to give a total size of 1\farcs1 (9.7 kpc). As seen in Figure \ref{disk_spec} (bottom), \ha~emission falls on the edge of bright OH sky emission (1.701 \micron), which may bias the kinematics by masking out \ha~emission that might be present at these wavelengths. However, there is a velocity gradient across the source of $\sim$300 \kms~which is fitted to the inclined disk model with a position angle of 65$^{\circ}$. In Figure \ref{twoobj}, we present \ha~flux and kinematics, and Figure \ref{diskfigs} illustrates the disk model which has a velocity offset of 103 \kms, plateau velocity of v$_{p}$=130 \kms~which is achieved in 0\farcs12 (1.0 kpc), and beam smearing corrected plateau velocities of v$_{p}$=195 \kms. Residuals of the disk fit are the highest of the three disk candidates with an average of 46 \kms. Dispersion does not peak between the \ha~flux distribution as would be expected within a two merger system. In contrast, dispersions are lower at the center of the \ha~emission ($\sigma$$\sim$90 \kms), but rise toward the edges in the lower S/N regions. We believe this galaxy is a candidate disk galaxy based upon its smooth velocity gradient, dispersions, and a good $\tilde{\chi}^{2}$=1.2 for the disk fit.

The stellar mass of Q2343-BM145 is the lowest of all of our sources with M$_{*}$ = 2.0x10$^{9}$ \msun, and has one of the highest gas fractions of 86\%. In contrast to our merger sources, the virial mass is in relative agreement with the total baryonic mass, and with enclosed mass estimated from dynamics of the disk fit. LRIS spectroscopy shows that Q2343-BM145 has prominent Ly$\alpha$ emission redshifted with respect to \ha~emission by 313 \kms. Interstellar absorption lines are blueshifted by $\Delta$v$_{abs}$=-175 \kms~with respect to \ha~emission. 

\section{Discussion}\label{discuss}

We have presented dynamics of six z$\sim$1.6 star forming galaxies, which all have similar SFR and spatial extents of $\sim$1\arcsec, but show distinctively unique dynamical characteristics of merging systems and rotating disks. Inclined disk models were well-fit to three sources with good $\tilde{\chi}^{2}$ and low residuals between the modeled and observed velocities of 20 to 46 \kms. We have investigated the effects of beam smearing on \ha~flux distributions, velocity offsets, and dispersions, and have shown that these disk candidates are distinguishable from merger candidates. We believe these sources are excellent candidates for rotationally supported disks at the observed epoch. In this section we ask whether these sources are precursors of present-day spiral galaxies, or are we catching these sources in a ``disk" phase that will later follow a different evolutionary path?

All three of our disk candidates achieve a maximum or plateau velocity within $\sim$1 kpc. These plateau radii are comparable to those of local spiral galaxies' thin disks which have values typically between 0.1 and 1 kpc \citep{sof99}. Plateau velocities derived from our inclined disk models are relatively low v$_{p}$$\sim$125 \kms~compared to local Sa and Sb galaxies, but are comparable to lower end Sc galaxy plateau velocities \citep{rubin85} and enclosed mass estimates \citep{meis83}. However, as we discuss in \S\ref{fluxmodel}, beam smearing may inherently underestimate these plateau velocities since our inclined-disk model does not take into account the intrinsic flux distribution of the source. If we assume our sources exhibit an exponential flux profile with a scale size comparable to the PSF core, then we find our plateau velocities are underestimated by $\sim$30\%. This would elevate our plateau velocities to $\overline{v}$$_{p}$=186 \kms, closer to the median of local Sc galaxies (175 \kms, \citealt{sof01}) and consistent with low end modern Sb galaxies. To grow to a median Sb velocity of 220 \kms would require 40\% additional enclosed mass which would likely allow the existing rotationally-supported disk to survive. This implies that if these disk systems are precursors to present-day spirals, then plateau radii and velocities only need to grow by modest amounts to reach typical current epoch masses.  

For the modern Milky Way, the dynamical mass using a highly flattened spheroid within the solar orbit  \citep{ghez08} is close to 4x10$^{11}$ \msun. If we assume that 70\% of that total mass lies in baryons and 30\% lies in dark matter \citep{binn98} and we remove the contribution of the bulge's mass (1.5x10$^{10}$ \msun: \citealt{fuhr04}), then 1x10$^{11}$ \msun resides in both the thin and thick disk at our solar galactic radius with 90\% of this mass contained in stars. The thick disk contains roughly 20\% of the total mass of the thin disk \citep{chab01, fuhr04}, which yields a total stellar mass of the Milky Way thin disk of M$_{*}$$\sim$2x10$^{11}$ \msun. Our observed z$\sim$1.6 disk candidates are quite compact with \ha~radii under $\sim$5 kpc. Scaling the Milky Way kinematics to 5 kpc leaves a stellar mass of 8x10$^{10}$ \msun at comparable volumes. In the z$\sim$1.6 galaxies the stellar masses we derive are a 0.2-1.3x10$^{10}$ \msun, however the baryonic mass in these systems ($\mu$$\sim$70\%) still largely remains in the form of gas. If we assume that these systems convert the majority of their gas fractions to stellar components then their final stellar mass would be comparable to the present-day Milky Way.

Our estimated halo masses for these disk candidates are of 10$^{12}$ \msun. Current theories of halo merger simulations show that at least 50\% of 10$^{12}$ \msun halos at z$\sim$1.5 will have a merger ratio larger than 1:3 \citep{stewart08}. However, statistics of merger halo rates currently poses a paradox for disk formation, since observationally at present-day epochs it is expected that 70\% of 10$^{12}$ h$^{-1}$ \msun dark matter halos contain disk dominated galaxies (\citealt{wein06, park07}, references therein). These high merger rates for 10$^{12}$  dark matter halos have led to recent simulations involving gas rich ($\mu$$\gtrsim$ 50 \%) merger driven scenarios for disk formation \citep{springel05, robertson06}, where disk systems with these high gas fractions are able to survive a large merger event. Given the observed fraction of disk galaxies at z$\sim$0 with a 10$^{12}$ \msun dark matter halo, our well-fit disk model with low residuals and set velocity plateau and radii, and our high gas fractions ($\mu$$\gtrsim$ 60 \%), we believe these z$\sim$1.6 disk candidates have an opportunity to persist to present-day spirals.

N-body simulations have demonstrated a scenario of ``inside-out"  disk formation, where lower angular momentum material first collapses to the center of the system and higher momentum material falls in later to be deposited at larger radii (eg, \citealt{bosch02, kauf07}). This scenario would increase the overall sizes of our disks, but would not significantly affect scale lengths and plateau velocities. In addition, disk formation models have grappled with not allowing in-falling gas to cool too rapidly, thus generating disks that are too small with steep velocity profiles. In order to inhibit contraction of baryonic material, some modelers have invoked energetic feedback mechanisms to heat the gas by supernova and star formation feedback in order to slow cooling \citep{thom05, governato07}. 

Recent IFS observations of z$\gtrsim$2 star forming galaxies find high observed dispersions which they suggest are primarily intrinsic to the sources. \citet{law07} LGS AO observations of three such sources had a mean observed dispersion of 86 \kms~and range of 72 to 101 \kms. A similar seeing-limited sample by \citet{schreib06} had a mean observed dispersion of 130 \kms with a larger range of dispersions from 70 to 240 \kms.  These studies have typically related the ratio of circular velocities to the inferred intrinsic dispersions (v$_{c}$/$\sigma_{\rm{int}}$) to determine whether these sources are dispersion-dominated. We have shown that taking beam smearing into account can explain the majority of  ``high" observed dispersions in our sample. Observed dispersions of two of our disk candidates, HDF-BMZ1299 and Q2343-BM145, could be explained by our level of beam smearing. We find that the ratio of circular velocity to our intrinsic dispersions for HDF-BMZ1299 and Q2343-BM145 are  v$_{c}$/$\sigma_{\rm{int}}$$\sim$15, which is comparable to the Milky Way and local spiral galaxies. In contrast, Q2343-BX344 has highly elevated dispersions, and even taking beam smearing into account leaves an estimated intrinsic dispersion of $\sigma$$_{\rm{int}}$=150 \kms and v$_{c}$/$\sigma$$\sim$1.  So in the higher redshift cases with their highest observed dispersions, and the one high dispersion disk within our sample, a source of intrinsic dispersion is needed. The source of intrinsic dispersion may be related to heating of the ISM by supernovae and star formation activity \citep{thom05, governato07}, or may simply reflect the dynamics of accreting material onto these disk systems. There is a selection bias towards compact sources with our LGS AO sample, while the seeing-limited sample was biased to detect more extended objects with larger dispersions, based on their selection criteria of galaxies that showed large velocity gradients and high dispersions with previous near-infrared long slit spectroscopy \citep{erb03}.

Two of our disk candidates have strong, concentrated \nii/\ha~ratios which indicate these systems are harboring AGNs. Both of these galaxies show slightly elevated 8.0$\micron$ flux compared to the other sources, but in general their photometric colors and rest-frame UV spectroscopy show no strong indications of having AGN activity. We thus believe the majority of the gas excitation is arising from ongoing star formation. Without the high spatial resolution and 2D spatial mapping from the OSIRIS-LGSAO system, it would be difficult to disentangle the weak contribution of AGN activity. This AGN emission would contaminate global metallicity measurements  (ie, \citealt{shap05, liu08}) and calls into question some estimates of seeing limited metallicity gradients at high redshift. Given the ages inferred from their SED fits ($\lesssim$ 1 Gyr), it is likely these disk candidates have only completed a few orbital periods since star formation began. This could help explain the relatively high dispersions and the asymmetries in \ha~with respect to the \nii~emission.  We may speculate on a simplified picture for the formation of these disk systems, where major gas infall and possibly a merger event occurs and produces a disturbed thickened disk with intense star formation. During the first few rotations dispersions may remain high as in Q2343-BX344, possibly due to feedback but also due to the original random motions of the gas. During that phase, angular momentum is being dissipated, creating a general infall including a small fraction of material into the central super-massive black hole. Eventually the AGN weakens, the disk stabilizes and star formation decreases, and these galaxies fall out of the classification of ``star forming galaxies". Both disk systems exhibit a similar \nii~redshift of $\gtrsim$100 \kms~with respect to \ha, suggesting a possible outflow from the AGN. Q2343-BX344 also has unusual characteristics of having interstellar absorption lines redshifted by $\sim$150 \kms, which suggests that gas is in-falling onto this system.

We note that near-infrared observations using LGS-AO and integral field spectroscopy are still in their technological infancy. Although we have presented results for only six galaxies, preliminary results of this on-going survey show that resolved spectroscopy of high-redshift galaxies offers unique insights into these systems. However, there is still a great deal of work needed to understand the observational complexities of beam smearing on the kinematics of these objects. Future research will include development of a large, mature survey of LGSAO IFS observations of high redshift galaxies, longer integrations on z$\sim$1.3 - 2.4 galaxies to achieve significant S/N for determining dynamics at larger radii, and using multiple nebular chemical indicators (ie, \oiii and \hb) for better determinations of metallicities and AGN contributions. Use of current facilities like OSIRIS and SINFONI will also help us plan and prepare for future large ground-based telescopes (ie, TMT, ELTs) with ever more sophisticated IFS instruments \citep{cramp08}.

\acknowledgements
The authors would like to acknowledge the dedicated members of the Keck Observatory staff, particularly Marcos van Dam, Jim Lyke, Randy Campbell, and Al Conrad,  who helped greatly with the success of our observations. Our referee offered interesting insight and beneficial suggestions. We would like to thank the generous funding from the Center of Adaptive Optics which supported this research program. Data presented herein were obtained at W.M. Keck Observatory, which is operated as a scientific partnership between the California Institute of Technology, the University of California and the National Aeronautics and Space Administration. The Observatory was made possible by generous financial support of the W.M. Keck Foundation. The authors wish to recognize and acknowledge the significant cultural role and reverence that the summit of Mauna Kea has always had within the indigenous Hawaiian community.  We are most fortunate to have the opportunity to conduct observations from this ``heiau" mountain.


\begin{deluxetable}{lcclcccccc}
\tabletypesize{\scriptsize}
\tablecaption{Observational Information \label{obs}}
\tablewidth{0pt}
\tablehead{
\colhead{Target} &
\colhead{RA} & 
\colhead{DEC} & 
\colhead{Date} &
\colhead{Exposure} & 
\colhead{PA} & 
\colhead{Filter} & 
\colhead{R$_{TT}$} &
\colhead{Sep$_{TT}$} & 
\colhead{Strehl\tablenotemark{a}}  \\ 
\colhead{Name} &
\colhead{(J2000)} & 
\colhead{(J2000)} & 
\colhead{Observed} & 
\colhead{(seconds)} &
\colhead{(deg)} & 
\colhead{Used} & 
\colhead{(mags)} & 
\colhead{($\arcsec$)} &
\colhead{(\%)} \\
}
\startdata
HDF-BX1287    & 12 36 20.6 & +62 16 57.6 &  2007 May 17 & 4x900     &  0  &Hn5 &14.5   & 34.5  &  30  \\
HDF-BX1315    & 12 36 30.1 & +62 16 35.4 &  2006 May 24  & 6x900    &   0 &Hn5 &14.5  &  36.1 &  30 \\
HDF-BMZ1299 & 12 37 01.3 & +62 08 44.3 &  2006 May 23  & 4x900    &  45  &Hn4 &16.6   &  55.3 &  22 \\
Q1623-BX491  & 16 25 54.2 & +26 43 52.4 &  2007 April 07 & 4x900     &  0  &Hn5 &15.3  &  11.5 &  29  \\
Q2343-BX344  & 23 46 26.5 & +12 47 07.2 &  2006 August 18 & 6x900 &  270 &Hn3 & 16.8  & 26.5  & 23  \\
Q2343-BM145 & 23 46 31.4 & +12 48 23.5 &   2006 August 19 & 5x900 & 270  &Hn4 &14.9 &  25.5 &  29 \\
\enddata
\tablenotetext{a}{Strehl ratio estimated for galaxy observations. This was estimated from the tip-tilt (TT) star OSIRIS-LGS performance and with Keck Strehl calculator given TT magnitude and separation.}
\end{deluxetable}

\begin{deluxetable}{lccccccccccccc}
\tabletypesize{\scriptsize}
\tablecaption{Photometric and SED Properties \label{photo}}
\tablewidth{0pt}
\tablehead{
\colhead{Name} & 
\colhead{U$_{n}$\tablenotemark{a}} & 
\colhead{G\tablenotemark{a}} &
\colhead{\it{R}\tablenotemark{a}} &
\colhead{J\tablenotemark{b}} &
\colhead{K$_{s}$\tablenotemark{b}} &
\colhead{m$_{3.6}$\tablenotemark{c}} & 
\colhead{m$_{4.5}$\tablenotemark{c}}  & 
\colhead{m$_{5.8}$\tablenotemark{c}}  &
\colhead{m$_{8.0}$\tablenotemark{c}} &
\colhead{F$_{24}$\tablenotemark{c}} &
\colhead{E(B-V)\tablenotemark{d}} &
\colhead{Age\tablenotemark{d}} &
\colhead{M$_{*}$\tablenotemark{d}} \\
\colhead{} &
\colhead{} &
\colhead{} &
\colhead{} &
\colhead{} &
\colhead{} &
\colhead{} &
\colhead{} &
\colhead{} &
\colhead{} &
\colhead{(Jy)} &
\colhead{} &
\colhead{(Myr)} &
\colhead{(10$^{10}$\msun)} \\
}
\startdata 
HDF-BX1287      & 23.44 & 23.10 & 23.05 & 21.27 & 20.76 & 22.60 & 22.50 & 22.68  & 23.27 & 35.1 & 0.28      & 9   & 0.2   \\
HDF-BX1315      & 24.37 & 23.96 & 23.77 & 21.72 & 20.24 & 21.79 & 21.75 & 21.91 & 22.05 & 64.8 & 0.22 & 570 & 1.5 \\
HDF-BMZ1299   & 24.26 & 23.87 & 23.31 & -          & -          & 20.91 & 21.04 & 21.10  & 21.27 & -         & 0.35 & 110 & 1.3 \\
Q1623-BX491    & 24.35 & 24.02 & 23.93 & 22.08  & 21.05  & 22.12 & 22.27 & 22.55  & - & - & 0.14 & 640 & 0.9  \\
Q2343-BX344    & 24.38 & 23.85 & 23.65 & 21.85 & 20.28 & - & 21.68 & - & 21.45  & - & 0.24 & 360 & 1.1 \\
Q2343-BM145   & 25.32 & 25.11  & 24.91 &  - & -  & -  & 23.34 & - & -  & - & 0.20 & 202 & 0.2 \\
\enddata
\tablenotetext{a}{Magnitudes are AB standard from \citet{erb06b}.}
\tablenotetext{b}{Magnitudes are Vega standard from \citet{erb06b}.}
\tablenotetext{c}{IRAC and MIPS magnitudes (AB standard) for HDF galaxies are from \citet{reddy06}; other galaxies IRAC and MIPS magnitudes are from private communication.}
\tablenotetext{d}{The typical uncertainty to the SED fits are  $\langle$$\sigma$$_{E(B-V)}$/E(B-V)$\rangle$ = 0.7,  $\langle$$\sigma$$_{Age}$/Age$\rangle$ = 0.5 , and $\langle$$\sigma$$_{M_{*}}$/M$_{*}$$\rangle$ = 0.4, and details of the SED fits are further described in \citet{erb06b}.}
\end{deluxetable}

\begin{deluxetable}{lccccccc}
\tabletypesize{\scriptsize}
\tablecaption{OSIRIS Global Properties \label{global}}
\tablewidth{0pt}
\tablehead{
\colhead{Name} & 
\colhead{z$_{H\alpha}$\tablenotemark{a}} & 
\colhead{F$_{H\alpha}$} &
\colhead{F$_{[NII]}$($\lambda$6585)} &
\colhead{\nii/\ha \tablenotemark{b}} & 
\colhead{12 + log(O/H)} & 
\colhead{SFR\tablenotemark{c}} & 
\colhead{SFR$_{dered}$} \\
\colhead{} &
\colhead{} &
\colhead{(ergs s$^{-1}$ cm$^{-2}$)} &
\colhead{(ergs s$^{-1}$ cm$^{-2}$)} &
\colhead{} &
\colhead{} &
\colhead{(M$_\odot$~yr$^{-1}$)} &
\colhead{(M$_\odot$~yr$^{-1}$)} \\
}
\startdata 
HDF-BX1287-N 	 & 1.6778 &  7.0$\pm$1.2x10$^{-17}$ &  - 					&  $\lesssim$0.3 &  $\lesssim$8.60 & 6$\pm$1 &  11$\pm$2 \\
HDF-BX1287-S  	& 1.6761  &  1.9$\pm$0.1x10$^{-16}$ & - 						&   $\lesssim$0.1 &  $\lesssim$8.33  & 16$\pm$1 & 31$\pm$2  \\
HDF-BX1315-NE      & 1.6735  &  3.3$\pm$0.7x10$^{-17}$  & - 					&  $\lesssim$0.2 &  $\lesssim$8.50 & 3$\pm$1 &  5$\pm$1  \\
HDF-BX1315-SW      & 1.6715  &  6.4$\pm$1.5x10$^{-17}$  & - 					&  $\lesssim$0.4 &  $\lesssim$8.67 & 6$\pm$1 &  9$\pm$2  \\
HDF-BMZ1299  	 & 1.5981  &  2.2$\pm$0.1x10$^{-16}$  &  6.3$\pm$0.8x10$^{-17}$  & 0.9$\pm$0.1 &  8.87 & 15$\pm$2 & 33$\pm$5  \\
Q1623-BX491   	 & 1.6446  &   1.7$\pm$0.3x10$^{-16}$  &  -  					& $\lesssim$0.2&  $\lesssim$8.50 & 14$\pm$2 & 19$\pm$3  \\
Q2343-BX344  	 	 & 1.5274  &   1.7$\pm$0.1x10$^{-16}$ & 6.9$\pm$0.8x10$^{-17}$ &  0.5$\pm$0.1 &  8.72 & 12$\pm$1 &  20$\pm$1 \\
Q2343-BM145   	&  1.5896 &   5.1$\pm$0.7x10$^{-17}$ & - 					&   $\lesssim$0.2 & $\lesssim$8.50  & 4$\pm$1 & 6$\pm$1  \\
\enddata
\tablenotetext{a}{Redshift is geocentric.}
\tablenotetext{b}{For measurements where there is a limit, the \nii/\ha~ratio was estimated by the predicted location of \nii~emission relative to the flux locations of \ha. For the two sources with direct \nii~detection, the ratio of \nii/\ha~was determined between the observed \nii~flux location to the \ha~emission.}
\tablenotetext{c}{Assuming concordance $\Lambda$-dominated cosmology \citep{benn03}, Kennicutt Law \citep{kenn98}, and \citet{chab03} IMF.}
\end{deluxetable}

\begin{deluxetable}{lcccccccc}
\tabletypesize{\scriptsize}
\tablecaption{\ha~Kinematic and Mass Distribution Properties \label{kinemass}}
\tablewidth{0pt}
\tablehead{
\colhead{Name} & 
\colhead{$\sigma_{global}$}  & 
\colhead{$\overline{v}$$_{error}$}  &
\colhead{r$_{1/2}$} &
\colhead{M$_{virial}$\tablenotemark{a}} &
\colhead{M$_{gas}$\tablenotemark{b}} &
\colhead{Gas Fraction} &
\colhead{M$_{enclosed}$\tablenotemark{c}} &
\colhead{M$_{halo}$\tablenotemark{c}}  \\
\colhead{} &
\colhead{(km s$^{-1}$)} &
\colhead{(km s$^{-1}$)} &
\colhead{(kpc)} &
\colhead{(10$^{10}$\msun)} &
\colhead{(10$^{10}$\msun)} &
\colhead{($\mu$)} &
\colhead{(10$^{10}$\msun)} &
\colhead{(10$^{12}$\msun)} \\
}
\startdata 
HDF-BX1287-N& 45$\pm$14 & 6 &3.9 & 0.6 & 0.3$\pm$0.1 & 0.95\tablenotemark{d} & - &  -  \\
HDF-BX1287-S    & 80$\pm$12 & 11 & 3.1& 1.4 & 4.6$\pm$1.6 & - & - &  -  \\
HDF-BX1315-NE  & 32$\pm$15 & 7 & 2.1&  0.2 & 1.0$\pm$1.0 & 0.60\tablenotemark{d} & - &   - \\
HDF-BX1315-SW     & 60$\pm$25 & 7 & 2.9 & 0.7 & 1.3$\pm$1.0 & - & - &   - \\
HDF-BMZ1299  & 79$\pm$2 & 10 & 2.5 & 1.1 & 4.4$\pm$1.1 & 0.77& 2.6 & 0.7 \\
Q1623-BX491     &  98$\pm$21 & 12 & 2.5 & 1.7 & 3.2$\pm$1.0 & 0.78 & - &   -  \\
Q2343-BX344       & 212$\pm$13 & 18 & 2.7 & 8.5 & 2.5$\pm$0.4 & 0.70 & 3.9 & 1.2   \\
Q2343-BM145     & 99$\pm$28 & 18 & 3.2 & 2.2 & 1.0$\pm$0.3 & 0.86 & 3.1 & 1.1   \\
\enddata
\tablenotetext{a}{For both the merger and disk candidates we assume a constant C=3 for the virial mass calculation.}
\tablenotetext{b}{Uncertainties associated with the gas mass (M$_{gas}$) are determined from the noise of a nearby sky region corresponding to the same spatial area of the source and wavelength range.}
\tablenotetext{c}{Plateau velocities (v$_{p}$) from disk models corrected for beam smearing were used.}
\tablenotetext{d}{Since M$_{stellar}$ was measured for the entire source, this gas fraction is represented for both components combined.}
\end{deluxetable}

\begin{deluxetable}{lcccccccc}
\tabletypesize{\scriptsize}
\tablecaption{Disk Model Parameters \label{diskmodel}}
\tablewidth{0pt}
\tablehead{
\colhead{Name} & 
\colhead{Inclination} &
\colhead{Position Angle}  &
\colhead{Slope} &
\colhead{v$_{o}$} &
\colhead{v$_{p}$} &
\colhead{r$_{p}$}  & 
\colhead{$\tilde{\chi}$$^{2}$} &
\colhead{Residual}  \\
\colhead{} &
\colhead{($\textit{i}$)} &
\colhead{($\phi$)} & 
\colhead{(km s$^{-1}$ kpc$^{-1}$)} &  
\colhead{(km s$^{-1}$)} &  
\colhead{(km s$^{-1}$)} &  
\colhead{(kpc)} &
\colhead{} &
\colhead{(km s$^{-1}$)} \\
}
\startdata 
HDF-BX1315      &   45      & 46 & 443  & 85 & 161 &0.5 & 3.4 & 30 \\
HDF-BMZ1299   &  45      &  315\tablenotemark{a}  & 205 & 25 & 111\tablenotemark{b} & 0.8  & 1.2 & 20 \\
Q2343-BX344    &   45     & 1 &  167 & 0 & 132\tablenotemark{b} &  1.1  & 0.4 & 27 \\
Q2343-BM145   &    45     & 55 & 179 & 103 & 130\tablenotemark{b} & 1.0  & 1.2 & 46\\
\enddata
\tablenotetext{a}{This model position angle ($\phi$) is measured with respect to an observed position (PA) angle of 45$^{\circ}$.}
\tablenotetext{b}{Correcting for beam smearing effects these plateau velocities (v$_{p}$) are estimated to be 30\% higher.}
\end{deluxetable}

\begin{deluxetable}{llccccc}
\tabletypesize{\scriptsize}
\tablecaption{Rest-Frame UV Spectral Observations and Properties \label{lris}}
\tablewidth{0pt}
\tablehead{
\colhead{Name} & 
\colhead{Date} &
\colhead{Exposure} & 
\colhead{Slit} &
\colhead{$\Delta$v$_{abs}$} &
\colhead{$\Delta$v$_{emi}$} &
\colhead{Emission} \\
\colhead{} &
\colhead{Observed} &
\colhead{(minutes)} &
\colhead{PA} &
\colhead{(km s$^{-1}$)} &  
\colhead{(km s$^{-1}$)} &
\colhead{Features} \\
}
\startdata 
HDF-BX1315         & 2005 May 06 			& 92 & -11 & -250  & -277 &  C{\sc iii}] \\
HDF-BMZ1299      & 2004 Apr 17 			& 245.3  & 0 & -354 &  - & - \\
Q2343-BX344       & 2003 Sep 24 / 2002 Oct 26 & 671 & 111& 144 & -122 & He{\sc ii}, C{\sc iii}]  \\
Q2343-BM145      & 2003 Sep 26 			& 330 & -112 & -175 & 314 & Ly$\alpha$, He{\sc ii}, C{\sc iii}]   \\
\enddata
\end{deluxetable}

\clearpage
\end{document}